\documentclass[12pt,a4paper]{article}
\pdfoutput=1
\usepackage{graphicx,amssymb,amsmath}
\usepackage[sort&compress,numbers]{natbib}

\graphicspath{{figs/}}
\usepackage[colorlinks=true,urlcolor=blue,
anchorcolor=blue,citecolor=blue,filecolor=blue,
linkcolor=blue,menucolor=blue,
linktocpage=true,pdfa=true]{hyperref}

\usepackage[left=2cm, right=2cm, top=1.9cm, bottom=0.75cm,
footskip=48pt, headsep=20pt, includehead, includefoot]{geometry}

\newcommand{\uhecrs}{\mbox{UHECRs}}
\newcommand{\uhecr}{\mbox{UHECR}}
\newcommand{\keuso}{\mbox{K-EUSO}}

\begin{document}
\title{\bfseries
	Status of the K-EUSO Orbital Detector\\ of
	Ultra-high Energy Cosmic Rays}

\author{Pavel~Klimov$^1$,
    Matteo~Battisti$^2$,
    Alexander~Belov$^{1,3}$,
    Mario~Bertaina$^2$,\\
    Marta~Bianciotto$^2$,
    Sylvie~Blin-Bondil$^{4,5}$,
    Marco~Casolino$^{6,7}$,\\
    Toshikazu~Ebisuzaki$^{6}$,
    Francesco~Fenu$^2$,
    Christer~Fuglesang$^{8}$,\\
    W{\l}odzimierz~Marsza{\l}$^9$,
    Andrii~Neronov$^{4,10}$,
    Etienne~Parizot$^4$,\\
    Piergiorgio~Picozza$^{6,7,11}$,
    Zbigniew~Plebaniak$^2$,
    Guillaume~Pr\'ev\^ot$^{4}$,\\
    Marika~Przybylak$^9$,
    Naoto~Sakaki$^6$,
    Sergey~Sharakin$^1$,\\
    Kenji~Shinozaki$^9$,
    Jacek~Szabelski$^9$,
    Yoshiyuki~Takizawa$^6$,\\
    Daniil~Trofimov$^{1,3}$,
    Ivan~Yashin$^1$ and
	 Mikhail~Zotov$^{1}$\thanks{Correspondence:
	 \texttt{zotov@eas.sinp.msu.ru}}\\[3mm]
    $^1$D.V. Skobeltsyn Institute of Nuclear Physics,\\
    M.V. Lomonosov Moscow State University, Russia\\
    $^2$Department of Physics, University of Torino, Turin, Italy\\
    %Via Pietro Giuria 1 10125,\\ 
    $^3$Faculty of Physics, M.V. Lomonosov Moscow State University, Russia\\
    $^4$Universit\'e de Paris, CNRS, Astroparticule et Cosmologie,\\
    F-75013 Paris, France\\
    $^5$Omega, Ecole Polytechnique, CNRS/IN2P3, Palaiseau, France\\
    $^6$RIKEN, Wako, Saitama, Japan\\
    $^7$INFN, Structure of Rome Tor Vergata, Rome, Italy\\
    $^{8}$KTH Royal Institute of Technology, Stockholm, Sweden\\
    $^{9}$National Centre for Nuclear Research, {\L }\'od\'z, Poland\\
    $^{10}$Laboratory of Astrophysics, Ecole Polytechnique Federale\\
	 de Lausanne, Switzerland\\
    $^{11}$Department of Physics, University of Rome Tor Vergata, \\Rome, Italy\\
}

\date{}
\maketitle

\begin{abstract}

	\keuso{} (KLYPVE-EUSO) is a planned orbital mission aimed at studying
	ultra-high energy cosmic rays (\uhecrs) by detecting fluorescence and
	Cherenkov light emitted by extensive air showers in the nocturnal
	atmosphere of Earth in the ultraviolet (UV) range.  The observatory
	is being developed within the \mbox{JEM-EUSO} collaboration and is
	planned to be deployed on the International Space Station after 2025
	and operated for at least two years.  The telescope, consisting of
	$\sim10^{5}$~independent pixels, will allow a spatial resolution of
	$\sim0.6$~km on the ground, and, from a 400~km altitude, it will
	achieve a large and full sky exposure to sample the highest energy
	range of the UHECR spectrum.  We provide a comprehensive review of
	the current status of the development of the K-EUSO experiment,
	paying special attention to its hardware parts and expected
	performance. We demonstrate how results of the K-EUSO mission can
	complement the achievements of the existing ground-based experiments
	and push forward the intriguing studies of ultra-high energy cosmic
	rays, as well as bring new knowledge about other phenomena
	manifesting themselves in the atmosphere in the UV range.

\end{abstract}

\tableofcontents

%______________________________________________________________________________
\section{Introduction}

It was back in 1961 that cosmic rays with energies
$\gtrsim$50~EeV---henceforth called ultra-high energy cosmic rays
(\uhecrs)---were first registered~\cite{Linsley1961}.  However, the
nature and origin of these most energetic particles in the Universe
still remains a puzzle. One of the main difficulties in UHECR studies is
their extremely low flux, of the order of a few particles per km$^2$~sr
per century.  This requires the construction of huge experimental
facilities able to collect sufficient statistics.  For example, the
array of surface detectors of the world's largest cosmic ray
observatory, the Pierre Auger Observatory (Auger), extends over an area
of 3000~km$^2$~\cite{PierreAuger:2015eyc}.  Surface detectors of the
second largest experiment, the Telescope Array (TA), occupy an area
around 700~km$^2$, and the array is being expanded to cover an area
approximately equal to that of Auger~\cite{TAx4-2021}.  Both experiments
employ a so-called hybrid technique for registering cosmic rays,
complementing surface detectors with stations of fluorescence telescopes
that observe ultraviolet emission of extensive air showers (EASs)
developing in the nocturnal atmosphere.  A great wealth of data has been
obtained in the two experiments through the years, but the nature and
origin of \uhecrs\ still remains an open question.  Furthermore, the
results of Auger and TA related to the energy spectrum of cosmic rays at
the highest energies show differences that could partly be ascribed to
systematic uncertainties between the two observatories, but might also
have a different origin, such as statistical fluctuations or even the
effect of a powerful source in the northern
hemisphere~\cite{Deligny:2020gzq, Abbasi:2021qO, PierreAuger:2020qqz,
2019ICRC...36..280H, 2019ICRC...36..482Y, TelescopeArray:2021gxg}. The
fact that the fields of view of Auger and TA only partially overlap
makes it difficult to cross-check the results to a full extent.

A revolutionary approach for studying \uhecrs\ was put forward in
1981~\cite{Benson-Linsley-1981}.  By that time, registering fluorescence
emission of EASs was already an established technique for studying
\uhecrs{}~\cite{Bunner1967, 1977ICRC....8..252M, 1983AIPC...96..191C}.
Benson and Linsley suggested placing a wide-field-of-view telescope with
a mirror 36~m in diameter and a $\sim$5000-pixel retina in a low-Earth
orbit (500--600~km), thus drastically increasing the exposure of the
experiment and providing an opportunity to study cosmic rays arriving
from the whole celestial sphere. Besides fluorescence, an orbital
telescope is able to register forward-beamed Cherenkov radiation
reflected from the Earth's surface or from dense cloud tops, providing
important additional information about the development of an EAS.  A
number of projects aiming to implement the idea have been suggested
since then: among them, Orbiting Wide-field Light collectors
(OWL)~\cite{OWL}, Tracking Ultraviolet Setup (TUS) and
KLYPVE~\cite{2001AIPC..566...57K, 2001ICRC....2..831A}, Joint Experiment
Missions for Extreme Universe Space Observatory JEM-EUSO (formerly
Extreme Universe Space Observatory on board the Japanese Experiment
Module)~\cite{JEM-EUSO}, Probe Of Extreme Multi-Messenger Astrophysics
(POEMMA)~\cite{POEMMA:2020ykm} and others.  TUS was the first one
launched into space on 28th April, 2016.  It was a telescope with a
Fresnel mirror of 2~m$^2$ and a focal surface consisting of 256 pixels
designed as a prototype of the KLYPVE project and  mostly aimed at
verifying the possibility of registering faint UV signals in the
nocturnal atmosphere with an orbital telescope~\cite{SSR2017}.

The TUS mission lasted from May, 2016, until the end of November, 2017.
Unfortunately, 20\% of the photomultiplier tubes (PMTs) comprising the
focal surface of the detector were destroyed during an accident on the
very first day of the experiment, and the sensitivities of the remaining
PMTs changed in comparison with pre-flight measurements. Nevertheless,
TUS registered a vast variety of signals taking place in the nocturnal
atmosphere of Earth in the UV range and demonstrated that a space
detector with the primary goal of searching for \uhecrs\ can be used as
a multi-disciplinary instrument with a wide science reach and with
unique sensitivity for those phenomena~\cite{JCAP2017,
2019RemS...11.2449K, nuclearites-ICRC2019, JCAP2020, Klimov:2021rxn}.

Mini-EUSO (Multiwavelength Imaging New Instrument for the Extreme
Universe Space Observatory or ``UV atmosphere'' in the Russian Space
Program) is a telescope operating on board the International Space
Station (ISS) in the ultraviolet (UV) range with a field of view of
$\simeq $44$^{\circ}$ and a ground resolution of
$\simeq6.3\times6.3$~km$^2$. It was launched with the uncrewed Soyuz
\mbox{MS-14} on 22th August, 2019. The first observations from the
nadir-facing UV-transparent window in the Russian Zvezda module took
place on 7th October, 2019. Since then, it has been taking data
periodically, with installations occurring every couple of weeks.  The
instrument is expected to operate for at least three years.  Mini-EUSO
has, so far, observed night UV emissions from the Earth of natural and
artificial origin, thousands of meteor candidates and several transient
luminous events---in particular, emissions of light and very low
frequency perturbation from an electromagnetic pulse (ELVEs), which are
observed as large ring-like upper atmospheric emissions that appear to
be expanding at superluminal speed---among other observations and
scientific objectives.  For more details on the detector and the data
gathered so far, see~\cite{CAPEL20182954, Mini-EUSO2021}.

\keuso{} (KLYPVE-EUSO) is one of the key projects of the JEM-EUSO
program~\cite{Bertaina:2021+i}.  It is the first mission in this
framework that will be able to detect \uhecrs\ from space.  The project
is included in the long-term program of experiments onboard the Russian
segment of the ISS. It was planned to be launched in 2024. However, due
to the complexity and high cost of the project, the possible launch date
was shifted to around 2026.  In what follows, we provide a comprehensive
review of the project status, design and scientific capabilities.  The
design presented below is a strongly modified version of what was shown
in~\cite{Panasyuk:20164y, Klimov2017ICRC}.  It is developed to satisfy
the size and weight constraints of the carrier and deployment
capabilities of the Russian segment of the ISS and to reach optimized
scientific goals.

%______________________________________________________________________________
\section{Scientific Goals}

The general scientific goals of the \keuso\ mission are similar to those
of other projects of the JEM-EUSO collaboration~\cite{Bertaina:2019+0,
Bertaina:2021+i}.  The ultimate purpose of the experiment is to make a
significant step towards revealing the nature and origin of \uhecrs, the
highest energy particles in the Universe. The key questions to answer
are: what are \uhecrs, what are their sources and what are their
acceleration mechanisms?

The most obvious way to investigate the origin of the cosmic rays is
the analysis of their arrival directions. A benefit of an orbital
experiment is that it can register cosmic rays arriving from the whole
celestial sphere. In case a telescope is placed on the International
Space Station, the exposure can be almost uniform, provided the mission
is sufficiently long~\cite{JE-exposure}.  This will make large-scale
anisotropy analysis both more statistically powerful and more reliable
by providing a larger significance for the same overall statistics and
by eliminating the problem of merging datasets from different parts of
the sky taken with different instruments, and therefore different
resolutions and different systematics~\cite{2014ApJ...790L..21A,
Auger-dipole-2017, diMatteo:20197z, Kim:2021mcf}.

The arrival direction distribution as measured by Auger is characterized
by a large scale dipole anisotropy pointing in the direction
$(233^\circ,-13^\circ)$ in the galactic
coordinates~\cite{Auger-dipole-2017}. Such a result is significant to
over five sigma levels, supporting the extragalactic origin of cosmic
rays above $8\times10^{18}$~eV. On the intermediate scales, both
TA~\cite{2014ApJ...790L..21A}  and Auger~\cite{2018ApJLett_starburst}
show indications of an anisotropy; in the case of Auger, with some
degree of correlation with classes of nearby sources. This makes it
mostly important to measure the \uhecr\ flux over the entire sky in a
consistent way, similar to how an instrument in space would generally
provide. \keuso\ would be able to make a major improvement in this
respect by measuring dipole and quadrupole moments in the arrival
direction distribution, which are affected by major uncertainties when
performed by different arrays on ground~\cite{Tinyakov:2021}. In
addition, \keuso\ will be able to test if there is a correlation between
the arrival directions of \uhecrs\ and the distribution of matter in the
nearby Universe.  It will also provide an opportunity to verify other
hypotheses related to the anisotropy of \uhecrs.  In particular, after a
year or two of operation in space, \keuso\ could provide sufficient data
to identify the signature of a nearby source in the large-scale UHECR
anisotropy~\cite{Kalashev:2019skq}.  Data will also provide some
information about the mass composition, providing directly comparable
measurements over the entire sky for the first time. In particular, even
though the mass sensitivity of \keuso\ is limited, thanks to the high
statistics of events, it will be possible to measure for the first time
the depth of the maximum above $\sim$50~EeV using the fluorescence
technique. Such a measurement is currently not feasible with
ground-based arrays due to the limited duty cycle of the fluorescence
telescopes.

Another key characteristic of \uhecrs\ is their energy spectrum, which
is of crucial importance for understanding their origin and acceleration
mechanisms.  A large amount of effort is being applied to such studies
at the leading ground-based experiments; see,
e.g.,~\cite{Deligny:2020gzq, Abbasi:2021qO}.  The spectra obtained at
Auger and TA demonstrate a good agreement at energies below 10~EeV,
except for a difference in the absolute energy scale, which is within
the systematic uncertainties of the experiments.  An even better
agreement is reached in the common declination band.  However, an
important difference between the spectra is observed at the highest
energies. Besides this, TA suggests different steepening positions for
events arriving below and above declination $\delta=24.8^\circ$, which
may indicate a different energy spectrum in the northern hemisphere.  On
the other hand, no indication of the declination dependence has been
found by Auger.  With its full-sky coverage and comparatively large
exposure, \keuso\ will provide data of the energy spectrum of \uhecrs\
arriving from the whole celestial sphere, therefore giving a chance to
clarify the reason for the different instances of steepening of the TA
spectrum in the two declination bands and the origin of the systematic
differences in the spectra of Auger and TA at energies above
$\sim$50~EeV in the common declination band.  We believe that these
results will address some of the open questions of \uhecr\ physics
outlined in~\cite{2017PTEP.2017lA101D,2019FrASS...6...23B}, especially
those related to anisotropy studies and the energy spectrum.

Supplementary tasks of the \keuso\ mission are studies of transient
luminous events in the atmosphere. In particular, as it was demonstrated
by the TUS experiment, an orbital telescope is a perfect tool for
observing ELVEs, which take place at heights of around 80--90~km.
\keuso\ will also be able to shed new light on puzzling bright flashes
registered by TUS far from thunderstorm regions and populated areas.

Besides this, the telescope will observe meteors, thereby providing
additional information for comprehensive understanding of the dynamics
of meteors in the Solar System.  It will also be able to register or
place new limits on the existence of nuclearites (hypothetical massive
strange quark matter nuggets); see~\cite{jemeuso-meteors,
nuclearites-ICRC2019, Piotrowski:2021a1, Paul:2021W0} for more details
and in-depth~discussions.

%______________________________________________________________________________
\section{The Detector}

K-EUSO is a mission led by the Russian Space Agency together with the
international JEM-EUSO collaboration to place an \uhecr\ observatory on
board the Russian segment of the ISS. The concept of the detector is
based on the mirror-type detector proposed at the Skobeltsyn Institute
of Nuclear Physics of Lomonosov Moscow State University (SINP MSU) in
2001~\cite{2001AIPC..566...57K}. It has passed several stages of
improvements since then to meet both scientific requirements and
technical feasibility. In 2010, the project was included in the
long-term program of experiments onboard the Russian segment of the ISS.
In 2012, SINP MSU finished the preliminary design stage of the \keuso\
telescope for UHECR measurements from the International Space Station.
It was designed as a large 10~m$^2$ mirror telescope with a focal
distance of 3~m and a field of view (FOV) of about
$\pm7.5^\circ$~\cite{Khrenov2004, Garipov2001}. However, it became clear
during the preliminary design phase that the parameters of the
instrument (observation area and image quality) do not allow one to
solve current problems in UHECR science due to a too small geometrical
exposure.  These considerations initiated the development of a new
optical system for the KLYPVE detector in order to increase the FOV and
to improve the spatial and angular resolution and the overall
performance of the instrument. This work is performed in close
cooperation with the JEM-EUSO collaboration since late 2013. To
eliminate the off-axis aberration, an additional corrective Fresnel lens
was introduced in front of the photodetector. Two versions of the
detector were developed: the Baseline and Multi-Eye Telescope System
(METS)~\cite{Panasyuk:20164y, Garipov2015BRAS}. These configurations
were later transformed to a Schmidt telescope
design~\cite{Klimov2017ICRC}.

Further feasibility studies demonstrated difficulties in the delivery
and installation of the instrument outside the ISS.  There were more
than 30 separate parts for delivery, and assembling the telescope
required many extra vehicular activities of astronauts. A decision was
made to return to the version of purely lens optics based on the
developments of the JEM-EUSO project~\cite{JEM-EUSO-Instrument2015ExA},
since it can provide optimal weight and size characteristics of
the~equipment.

\begin{figure}[!ht]
	\centering 
	\includegraphics[height=5cm]{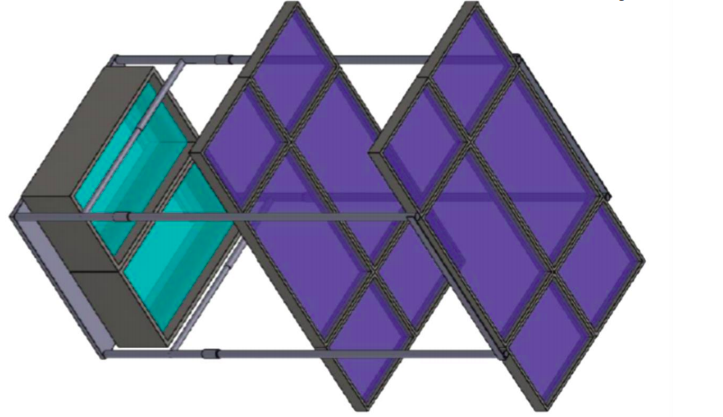}
	\includegraphics[height=5cm]{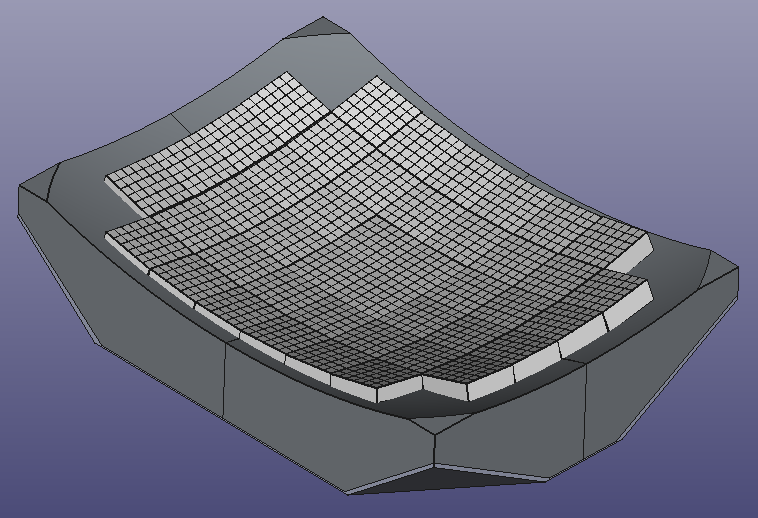}
	
	\caption{Left: a scheme of the \keuso{} detector with segmented
	refractive optics and rectangular aperture as designed by GP Advanced
	Project company. Right: a 3D model of the \keuso{} focal surface
	built from 44 photo detector modules. It is to be placed inside the
	green ``box'' shown in the left panel.}
	
	\label{fig:scheme}
\end{figure}

In the latest configuration presented below, the detector consists of a
refractive optical system with a rectangular aperture of
$1400~\text{mm}\times2400~\text{mm}$; see Figure~\ref{fig:scheme}. This
size is due to the fact that each frame in the folded state should be
smaller than $1200\times700\times350$~mm$^3$ and freely pass through the
Progress cargo hatch.  The optics comprises two optical elements
(lenses) that focus the light onto a focal surface of
$1300~\text{mm}\times1000~\text{mm}$ size.  The focal surface structure
consists of 44 photodetector modules (PDMs) similar to those of the
other JEM-EUSO missions, with a total amount of channels close to
10$^5$.  The main parameters of the detector are given in
Table~\ref{tab:KEUSOparameters}.  Spatial and temporal resolutions are,
respectively, given by the ground-projected area observed by an
individual channel (pixel) and the electronics gate time unit (GTU).
More details about the detector substructures are provided in the
following sections.

\begin{table}[!ht]
    \centering
    \caption{Technical parameters of \keuso}

    \smallskip
    \begin{tabular}{|l|c|}
        \hline
        Altitude & $\approx 400$~km\\
        \hline
		  Field of View & 0.3 sr (48 000 km$^2$)\\
        \hline
        Pixel size & $3~\text{mm}\times3~\text{mm}$ \\
        \hline
        Spatial/Temporal Resolution & 0.6 km (pixel) / 2.5~$\mu$s (GTU) \\
        \hline
        Entrance Pupil Area & $\sim$3~m$^2$ \\
        \hline
        Number of PDMs/channels & 44/101,376 \\
        \hline

        Dimensions (unfolded) & $\approx 125\times 250\times400$~cm$^3$\\
        \hline
    \end{tabular}

    \label{tab:KEUSOparameters}
\end{table}

\subsection{Optical System}

Both optical elements of the telescope are made in the form of spherical
Fresnel lenses manufactured from a special near-UV-transparent version of
poly methyl methacrylate (PMMA) material.
A significant optical power of such lenses is created by special
annular zones of the Fresnel
surface\footnote{\url{https://en.wikipedia.org/wiki/Fresnel_lens}
(accessed on 17 January 2022).
},
while, even with large dimensions, the thickness of the lenses turns out
to be small, and the mass of the entire system is moderate.
The PMMA-000 (Mitsubishi Chemical Corporation product) was chosen for
several reasons. Along with a low density (approximately 1.2~g/cm$^3$), which
ensures a low weight of lens, this material has a number of important
optical properties, including a weak temperature dependence of the
refractive index.  The transparency of PMMA-000 practically does not
change upon prolonged exposure to atomic oxygen (the main element of the
residual atmosphere at the ISS altitude) and protons with an energy of
approximately 70~MeV. At the same time, PMMA-000 is hard enough (in comparison,
for example, to soft and plastic CYTOP) and can be used for high
precision milling.  To date, a number of lenses with similar
characteristics of Fresnel concentric structures have already been
manufactured: lenses for balloon experiments
EUSO-Balloon~\cite{Adams2015-2} and EUSO-SPB1~\cite{Wiencke2017}, as
well as for the space experiment Mini-EUSO~\cite{CAPEL20182954}.

The front lens is convex-concave, the rear one is concave-convex and
all four radii of curvature are equal in absolute value, which greatly
simplifies their production. 

The optical power of the lenses is created by radial Fresnel structures,
which are located on both surfaces of the front lens (S1 and S2) and on
the front surface of the rear (S3). The calculation showed that, in order to
ensure the quality of signal focusing in a wide spectral range (from 300
to 400~nm), the remaining surface (S4) must be a diffractive optical
element~(DOE).

The selection of the optimal values of the parameters of all four
surfaces and the distances between the elements (lenses and focal
surface) was carried out within the framework of numerical simulation in
the Zemax
OpticStudio\footnote{\url{https://www.zemax.com/pages/opticstudio}
{(accessed on 13 December 2021).}
}.
Fresnel structures  were modeled using the Zemax extended Fresnel
surface with asphericity parameters up to the fourth degree in the square
of the radial distance $\rho^2$.  The DOE was simulated as a binary
optics~2 surface using the polynomial phase function up to the fifth
degree on $\rho^2$.

At the stage of preliminary modeling, the value of the radii of
the curvature of all optical surfaces $R_1 = R_2 = -R_3 = -R_4 = 5500$~mm,
as well as the concave spherical focal surface, $R_\mathrm{FS} =
2000$~mm, was chosen. The thickness of both lenses was fixed at 10~mm.

\begin{figure}[!ht]
	\centering 
	\includegraphics[height=5cm]{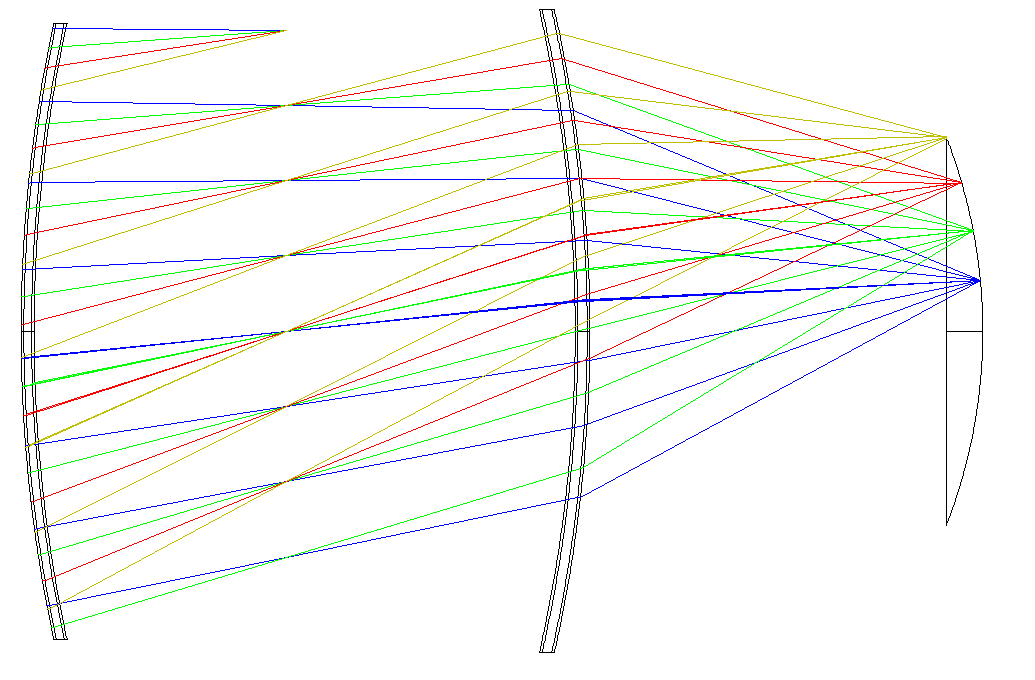}
	\includegraphics[height=5cm]{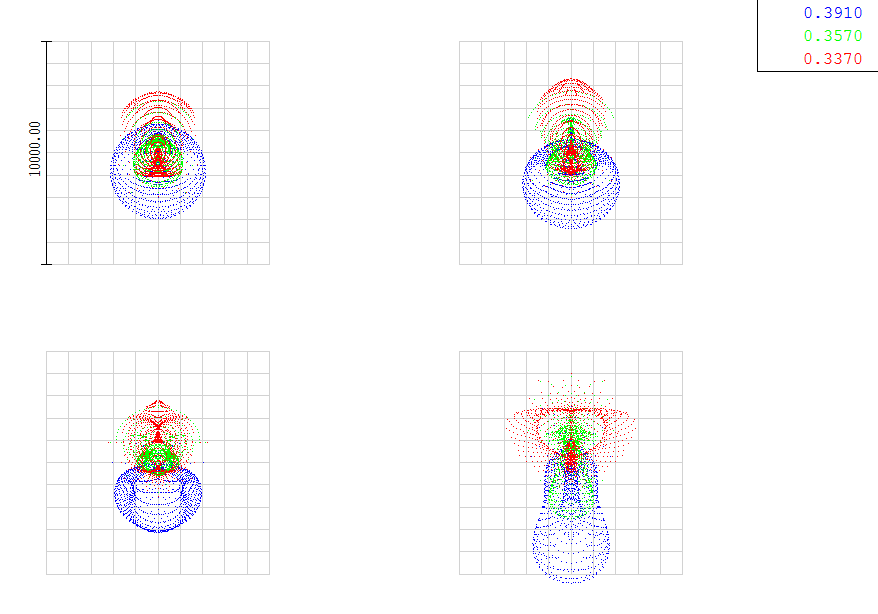}

	\caption{Left: K-EUSO optical scheme as sequential ray tracing (SRT)
	model in Zemax.  Right: spot diagrams for $\gamma=5^\circ,
	10^\circ,15^\circ,20^\circ$.  The grid cell has a size of
	$1~\mathrm{mm}\times1~\mathrm{mm}$.}

	\label{fig:optic_scheme}
\end{figure}

The RMS radius of the polychrome image (with the same weights for three
main wavelengths, 337, 357 and 391~nm), averaged over the field of view
from $0^\circ$ to $20^\circ$, was chosen as an optimization criterion.
The optical scheme and spot diagrams of the resulting system are shown
in Figure~\ref{fig:optic_scheme}.

The system has a rectangular aperture of
$1400~\text{mm}\times2400~\text{mm}$ and the following axial distances:
the distance between the lenses S2--S3 equals 2102~mm and the distance
between the rear lens and the focal surface S4--FS equals 1466~mm.  The
total axial length of the  system is $\text{S1}-\text{FS} = 3588$~mm;
thus, the longitudinal dimension of the entire telescope, including the
structure of the focal surface, does not exceed 4~m.  The characteristic
period of the diffraction grating at a distance of 1000~mm from the axis
is 6~$\mu$m and is acceptable from the point of view of manufacturing
technology.

The Zemax sequential ray tracing (SRT) model provides an approximate
estimate of the performance of the optical system. For a more accurate
analysis that takes into account both scattering by Fresnel structures
of optical surfaces and rectangular apertures of elements, the so-called
non-sequential (NSRT) model is required. In this model, the Fresnel
surfaces were chosen as annular zones of an equal depth of 2~mm. The total
numbers of zones are 329 (S1), 391 (S2) and 364 (S3).  To calculate the
detector response, it is necessary to implement it within the simulator
of the entire space experiment. Such models have been developed by the
\mbox{JEM-EUSO} collaboration in the framework of the ESAF package (EUSO
Simulation and Analysis Framework~\cite{Berat2010,Fenu:2019+0}).  As a
part of the calculations in ESAF, the consistency of the optics and
electronics of the K-EUSO was checked (i.e., the comparison of the image
size with the pixel size), as well as the calculation of the focusing
efficiency and the effective light collection area (throughput). 

\begin{figure}[!ht]
	\centering
	\includegraphics[width=4cm]{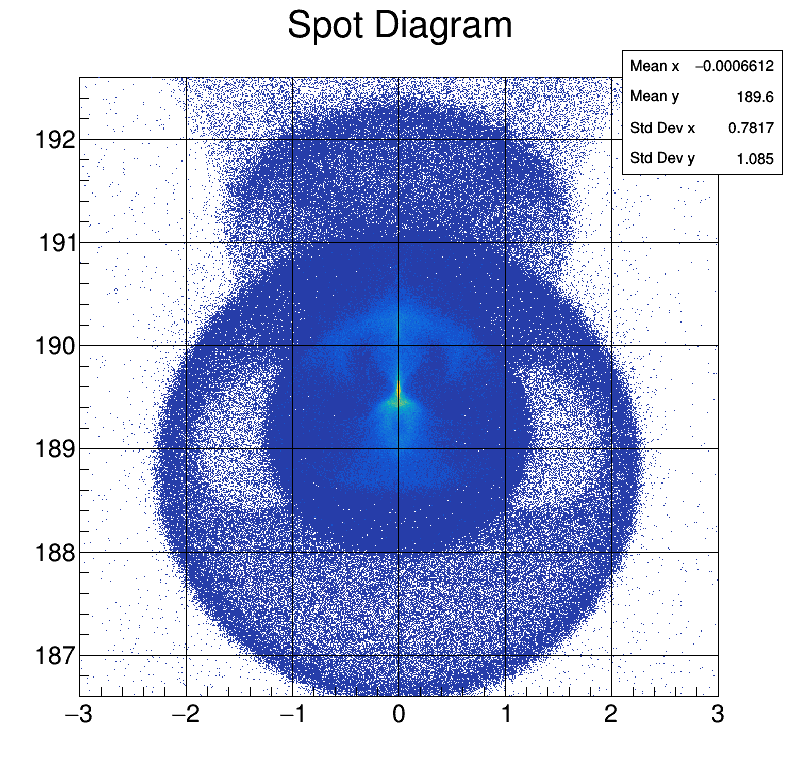}
	\includegraphics[width=4cm]{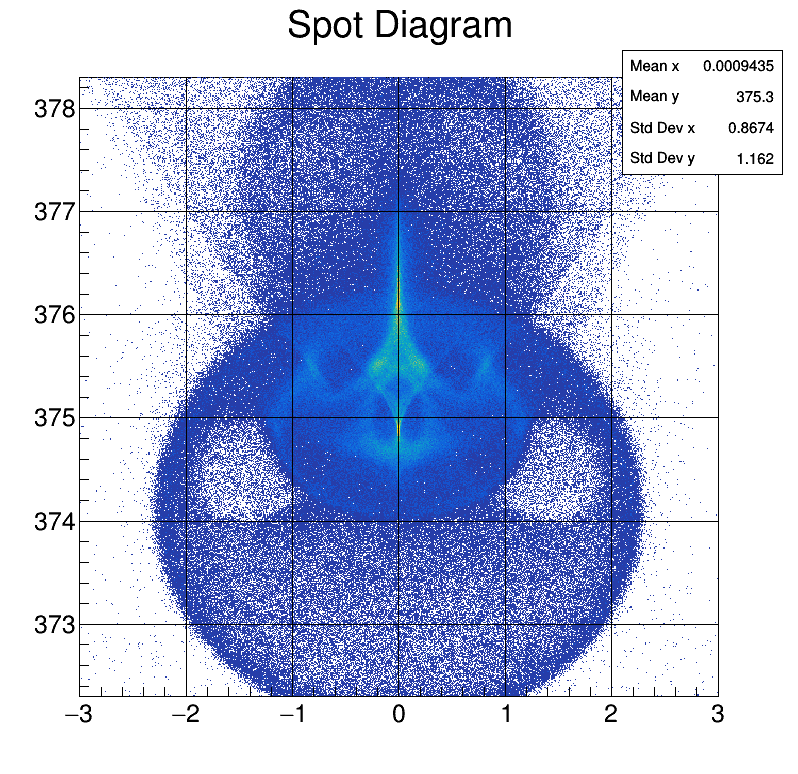}
	\includegraphics[width=4cm]{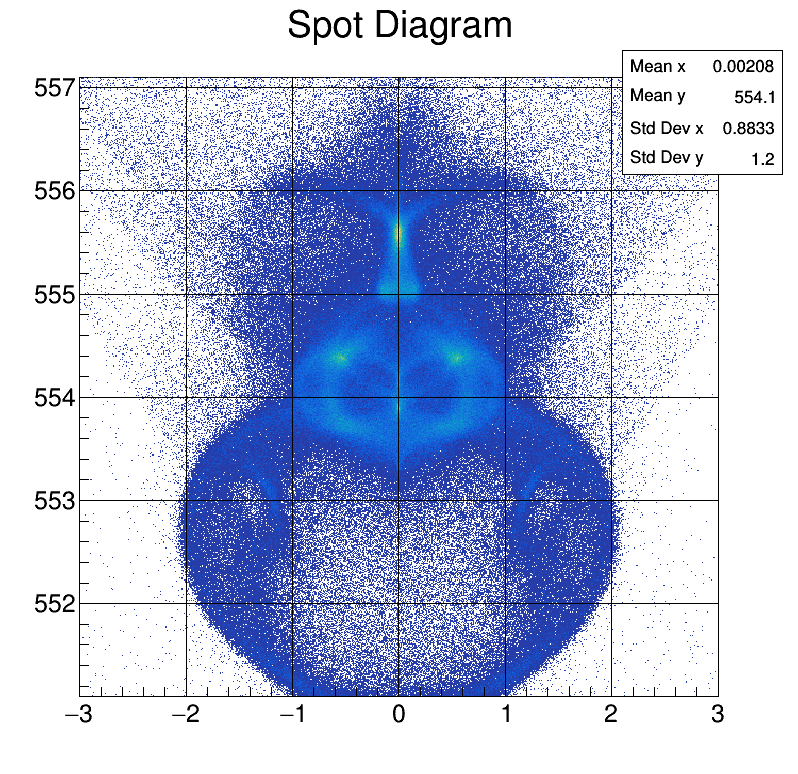}
	\includegraphics[width=4cm]{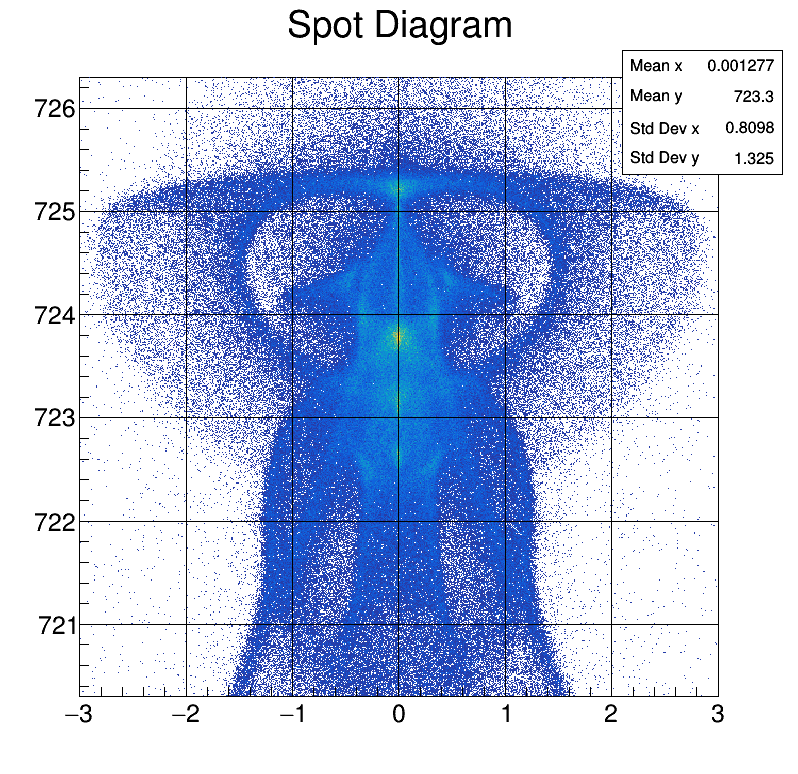}

	\caption{Spot diagrams of the NSRT model for $\gamma=5^\circ,
	10^\circ, 15^\circ, 20^\circ$. Grid cell has a size of
	$1~\mathrm{mm}\times1~\mathrm{mm}$.}

	\label{fig:SpotDiagrams}
\end{figure}

A characteristic feature of Fresnel optics is that a part of the light
energy is scattered at large angles relative to the centroid of the
image. By the image, we mean here the part of it that falls into a box
of size $6~\text{mm}\times6~\text{mm}$ (i.e., $2\times2$ \keuso\
pixels).  Spot diagrams are presented in Figure~\ref{fig:SpotDiagrams} for
different field angles~$\gamma$.  The quantitative dependence of the
image size, understood as the RMS diameter in the spot box, in the
entire field of view is presented in the third row of
Table~\ref{tab:OpticsPerformance} (here, $\phi$ is the azimuth angle,
$\phi = 0^\circ$ and $90^\circ$ along small and large sides of the
aperture, respectively).  It can be seen that the optical system matches
with a photodetector with a pixel of size $a = 3$~mm up to $\gamma =
15^\circ$. At larger angles, the asymmetry of the image becomes
significant, which leads to a decrease in the efficiency of light
collection. 

\begin{table}[!ht]
    \centering
    \caption{K-EUSO optics performance: image size $d_\mathrm{RMS}$, effective
    area~$S_\mathrm{eff}$ and ensquared energy EE as a function of field
    angles~$\gamma$,~$\phi$.}
    \smallskip
    \begin{tabular}{|c|*{8}{c|}}
        \hline
        $\gamma$ & \multicolumn{2}{c|}{$5^\circ$} & \multicolumn{2}{c|}{$10^\circ$} & \multicolumn{2}{c|}{$15^\circ$} & \multicolumn{2}{c|}{$20^\circ$}\\
        \hline
        $\phi$ & $0^\circ$ & $90^\circ$ & $0^\circ$ & $90^\circ$ & $0^\circ$ & $90^\circ$ & $0^\circ$ & $90^\circ$ \\
        \hline
        \hline
        $d_\mathrm{RMS}$, mm & 2.61 & 2.68 & 2.84 & 2.95 & 3.11 & 3.10 & 3.86 & 3.32 \\
        \hline
        $S_\mathrm{eff}$, m$^2$ &  1.53 & 1.49 & 1.30 & 1.34 & 0.98 & 1.22 & 0.56 & 0.94 \\
        \hline
        EE & 0.79 & 0.78 & 0.76 & 0.74 & 0.71 & 0.71 & 0.47 & 0.65 \\
        \hline
    \end{tabular}
    
    \label{tab:OpticsPerformance}
\end{table}

We estimated the efficiency of the optical system as the effective area
$S_\mathrm{eff}$, i.e., a ratio of the radiation energy in the image to
the illumination of the entrance pupil, as well as ensquared energy (EE),
the ratio of the energy in the pixel to that in the whole image.  When
calculating them, the effects of PMMA absorption, the reflection from
four optical surfaces (including total internal reflection) and the
scattering of rays due to hitting the lateral (cylindrical) sections of
the Fresnel grooves were taken into account, and the diffraction
efficiency of DOE was assumed to be 80\%.  The results are presented in
the last two rows of Table~\ref{tab:OpticsPerformance}. 

Azimuthal asymmetry practically does not affect the position of the
image: with a high degree of accuracy, the radial distance to the
centroid $\rho_\mathrm{c}$ is proportional to~$\gamma$ in the entire FOV
up to 20$^\circ$, and the effective focal length is
$F=\Delta\rho_\mathrm{c}/\Delta\gamma = 2070$~mm. 

Based on the obtained results, it can be argued that this optical system
of the telescope has the following characteristics:

\begin{itemize}

	 \item Field of view (with the optics matched to the sensor size):
		 asymmetric with maximum field angle $\gamma_{\max}=
		 18^\circ\text{--}20^\circ$ at $\phi=90^\circ$,
		 $\gamma_{\max}=15^\circ\text{--}16^\circ$ at $\phi=0^\circ$,
		 overall $\Omega\approx 0.3$~sr;

	 \item Resolution: angular $\Delta\gamma = a/F = 1.5$~mrad $\approx
		 0.1^\circ$, spatial $\Delta L = R\Delta\gamma = 0.6$~km (at the
		 orbit height $R=400$~km);

	 \item Light collection area: geometric (at entrance pupil)
		 $S_\mathrm{geom}= 3$~m$^2$, effective $S_\mathrm{eff}=
		 1.0\text{--}1.5$~m$^2$.

\end{itemize}

\begin{figure}[!ht]
	\centering
	\includegraphics[height=5cm]{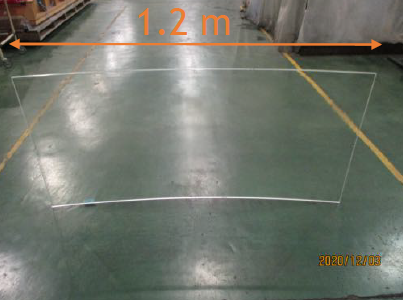}
	\includegraphics[height=5cm]{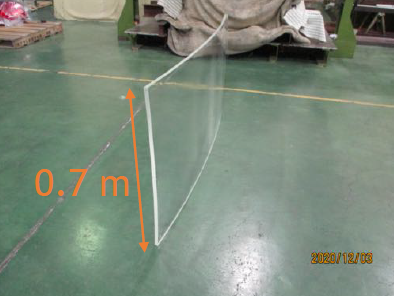}
	\caption{PMMA samples after preliminary shaping.}
	\label{fig:PMMA_samples}
\end{figure}

The manufacturing of the first (front) telescope lens began in late 2020.
The calculated radial width of the Fresnel zones varies from 16~cm in
the center to 1~mm at the periphery, with a groove depth of 2~mm.  In
view of the complexity of creating such a single one-piece large-sized
structure, it was decided to arrange each of the lenses from six
rectangular segments: four small ones, $700~\text{mm}\times600~\text{mm}$
and two large ones, $1200~\text{mm}\times700~\text{mm}$.  Each of the
segments were first given a spherical shape using two molds with a radius
of curvature slightly larger than the calculated one; see
Figure~\ref{fig:PMMA_samples}. 

\begin{figure}[!ht]
	\centering
	\includegraphics[height=4cm]{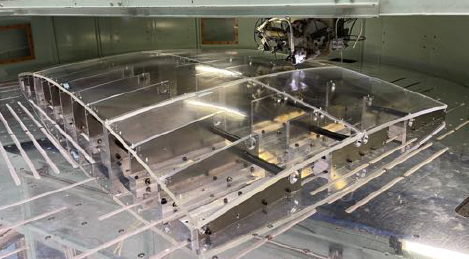}
	\includegraphics[height=4cm]{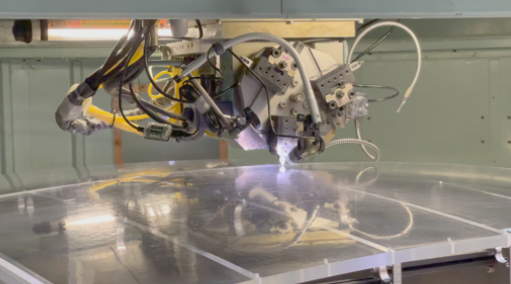}

	\caption{Lens shaping as a spherical surface of calculated curvature.
	On the left photo one can see special jig construction in order to
	assemble the lens from segments.}

	\label{fig:LensFormation}
\end{figure}

\begin{figure}[!ht]
	\centering
	\includegraphics[height=6cm]{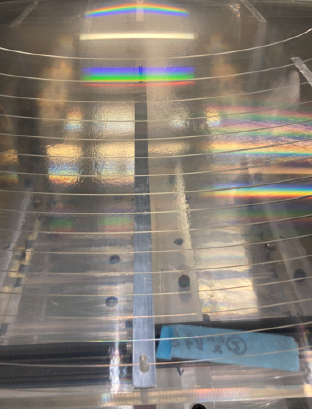}
	\includegraphics[height=6cm]{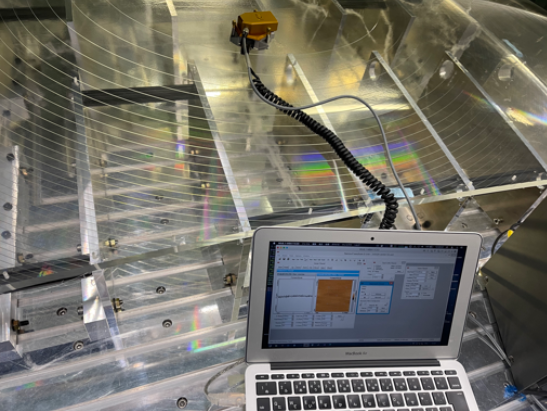}
	\caption{Formation and test of the Fresnel structure.
	Left: the central segment of the lens.
	Right: the measuring phase of the surface roughness.}
	\label{fig:FresnelSurfaceManufacturing}
\end{figure}

The further manufacturing process included imparting the spherical lens
to the calculated curvature, forming a radial Fresnel structure and
polishing it. For this, a special fastening structure was realized, which
made it possible to firmly attach curved plastic samples to it; see
Figure~\ref{fig:LensFormation}.   A Shibaura high-precision turning machine
UTD3400\footnote{\url{https://www.shibaura-machine.co.jp/en/product/nano/lineup/utd/shiyo.html}
(accessed on 13~December 2021).}
from RIKEN was used to make the Fresnel surfaces.
At the first stage, the samples fixed to the structure were given a
high-precision spherical shape with a calculated radius of curvature $R
= 5500$~mm. Then, a radial Fresnel structure with a depth of 2~mm was
applied to the surface (Figure~\ref{fig:FresnelSurfaceManufacturing}) and
the roughness was monitored. Measurements with Nanosurf Easyscan~2 AFM
(an atomic force microscope) showed root mean square surface roughness
$S_\mathrm{q} = 22$~nm.

The rear lens manufacturing will be finished by the end of March
2022, with the same quality as that of the front lens.  The performance
of the optics will be verified in~2022.  A special laser scan system was
manufactured at RIKEN for testing the quality of lenses.  It was used
earlier for testing the quality of the EUSO-SPB2 corrector
lens~\cite{2021EUSO-SPB2_OptTest}.  We expect that the optics performance
will meet the requirements.

\subsection{Photodetector and the Central Processor Unit}

The focal surface (FS) of the telescope is composed of 44 photodetector
modules. The structure of the PDMs is similar to that of the JEM-EUSO
telescope~\cite{JEM-EUSO-Instrument2015ExA} and employs a modification
of the digital data processing system (DDPS) developed previously for
the Mini-EUSO mission~\cite{CAPEL20182954}. The focal surface is a
spherical concave surface with a 2~m radius of curvature, but
constructed by flat PDMs.  The structure of the FS is shown in the right
panel of Figure~\ref{fig:scheme}.

Each PDM is a matrix of 36 Hamamatsu R11265-103-M64 Multi Anode
Photomultipliers (MAPMTs). Each MAPMT consists of 64 independent
channels (8 per side) with a 3~mm size.  Each of these channels
(identified as pixels in the following) has a field of view of
0.1$^\circ$, which corresponds to $\sim$600~m on the ground.  The gate
time unit (GTU) can vary from 1~$\mu$s to 2.5~$\mu$s, and will be
determined as a trade-off between the limited hardware and telemetry
budgets and the need for a good time resolution. Four MAPMTs are combined
in elementary cells (EC-units) with a common high voltage
Cockroft--Walton power supply.

The quantum efficiency of the MAPMTs is between 35--40\% in the
wavelength range of 300--400~nm.  The photomultiplier signal is read out
and amplified by the SPACIROC3 ASIC~\cite{blin2014}. The SPACIROC3
operates in the single photoelectron mode and has a double pulse
resolution of less than 10~ns. The majority of the data handling tasks,
such as data buffering, configuration of the read-out ASICS, triggering,
synchronization and interfacing with the central processor unit (CPU)
system, is performed by the DDPS system.  Given the very high time
resolution of the detector ($\sim$1~MHz) and the high number of pixels,
no full data retrieval is possible.  Data must therefore satisfy strict
trigger conditions.  Concentrations of the signal are sought for by the
trigger algorithms to preferentially select the shower signal while
rejecting background events; see Section~\ref{sec:performance}.

The DDPS consists of one Zynq-board and three Cross boards. The output
of the 36~ASICs of a PDM is collected by three Artix~7 FPGA-based Cross
boards. The Cross boards perform data gathering from the ASICs and data
multiplexing. The three Cross boards are connected to a Zynq board,
containing a Xilinx Zynq 7000 FPGA with an embedded dual core ARM9 CPU
processing system. The Zynq board controls the data flow from the ASICs,
runs the trigger logics and interfaces with other PDMs and the CPU for
data storage.

All PDMs of the FS operate independently and a special version of the
modular photodetector and digital electronics for on-board signal
processing are based on the network architecture principle. The network
principle is implemented by organizing three types of links: high-speed
communication between adjacent photodetector modules, long-haul
communication for recording information in the CPU ROM and synchronizing
communication for timing the operation of individual
modules~\cite{Belov2018}. Digital processing, including the trigger
system, is performed in a Zynq system-on-chip that includes the FPGA and
the processor system. The two-level trigger system for EAS detection is
used as was designed for the JEM-EUSO. For the registration of slower
events, slower modes of operation are foreseen to be similar to the
Mini-EUSO telescope. Triggers of the slow modes can be implemented in a
processor part of FPGA since they do not need to be extremely~fast.  

\begin{figure}[!ht]
   \centering
	\includegraphics[width=.42\textwidth]{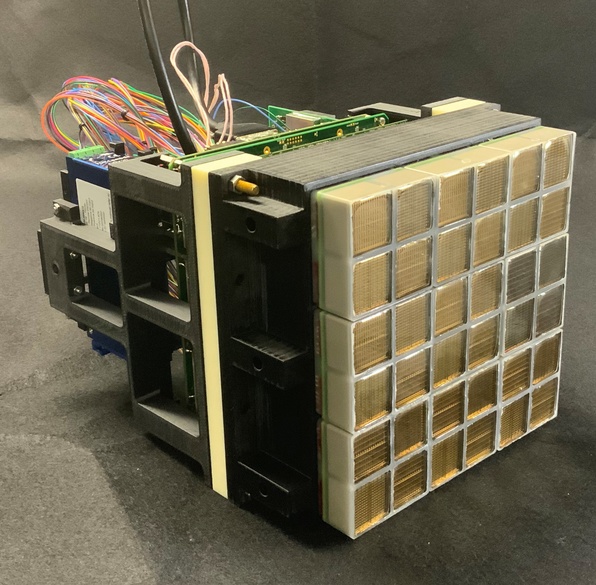}
	\caption{\keuso{} electronics. Photo of an assembled PDM.
	}
	\label{fig:FS}
\end{figure}

A model of the PDM was made during the preliminary design stage of the
project to confirm its performance characteristics and to check command
and data transfer with the CPU. A photo of the PDM with DDPS and nine EC
units during tests is shown in Figure~\ref{fig:FS}. The modified EC unit
with ASIC inside was developed especially for \keuso\ and EUSO-SPB2
projects in the Astroparticle and Cosmology laboratory (APC), France.

\begin{figure}[!ht]
	\centering
	\includegraphics[width=.49\textwidth]{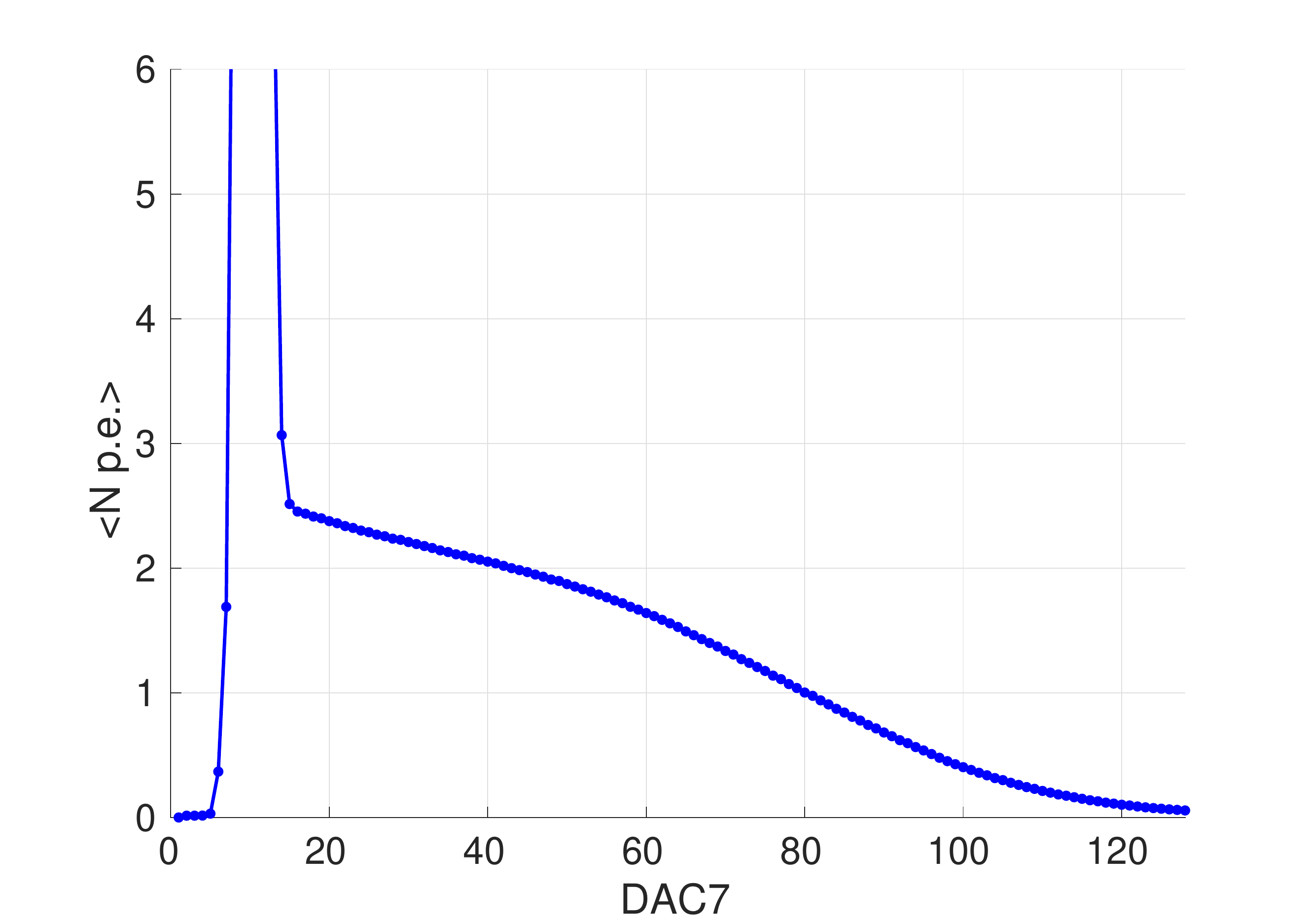}
	\includegraphics[width=.49\textwidth]{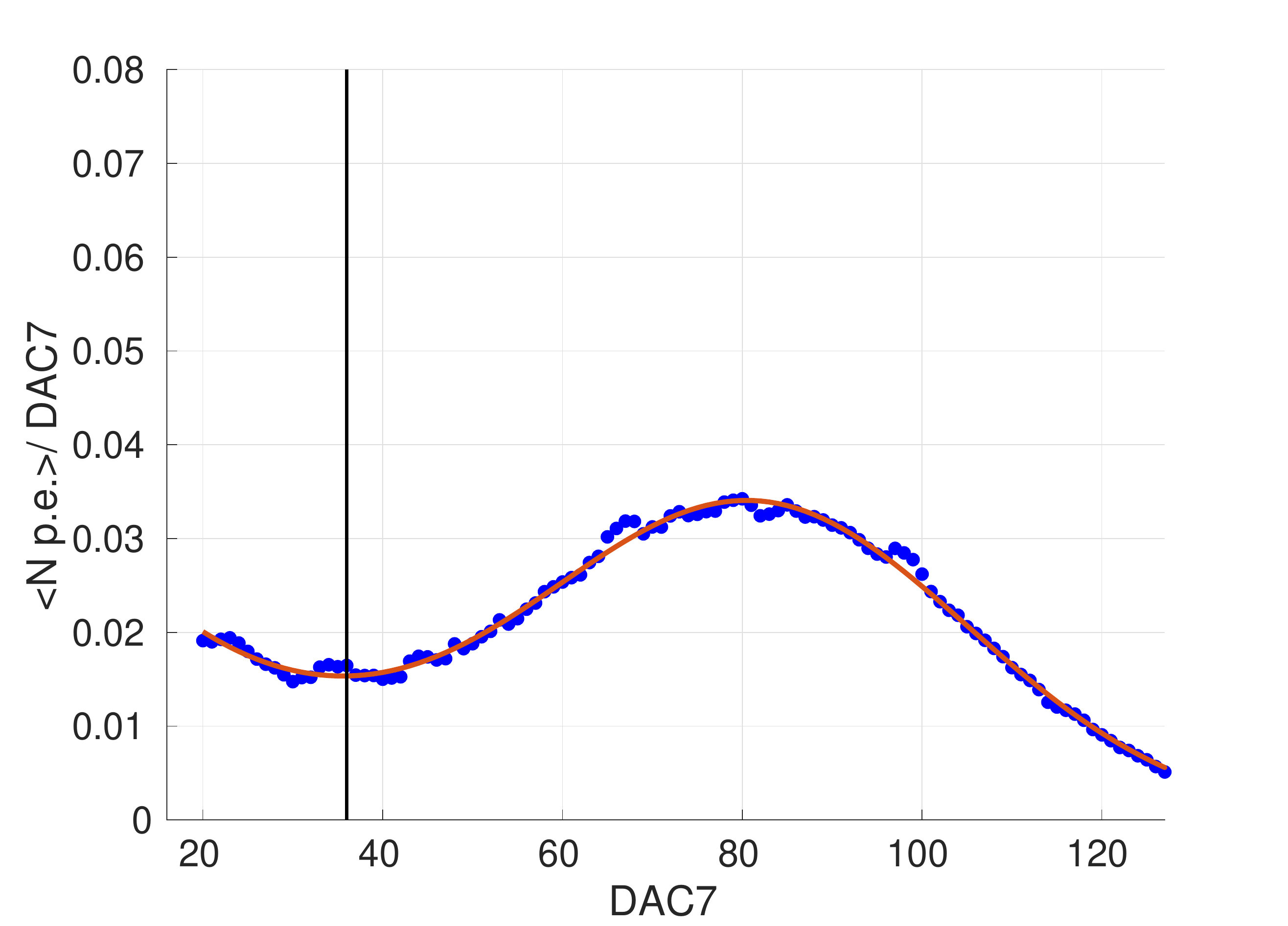}

	\caption{Left: S-curve of one K-EUSO photodetector pixel. Right: Single
	photoelectron spectrum for the same pixel. Blue dots show experimental data,
	the red line is their fit, the black vertical line shows the optimal
	threshold.  $\langle N_\mathrm{p.e.}\rangle$ is the mean number of
	photoelectrons per GTU, DAC7 is a threshold set in SPACIROC for pulse
	discrimination.}

	\label{fig:S-curve}
\end{figure}

The absolute photometric calibration of the PDM was performed to study
the overall efficiency of the photon detection.  The photometric
calibration provides a relationship between the amount of photons
arriving at the detector and the measured signal. The methods suggested
in~\cite{Caibration2015ExA} were employed for this purpose.  At the
first stage, the S-curves of all pixels were obtained to determine an
optimal threshold of a single pulse detection.\footnote{An S-curve is a
dependence of the ASIC count rate vs.\ threshold, so it is a cumulative
distribution function of single photoelectron~pulses.}
In \keuso, it is possible to adjust this threshold (the so-called DAC7
value) for each channel individually. A threshold is set in a minimum of
a valley of a single photoelectron pulse distribution (between real
photons and electronics noise).  An example of an S-curve for one K-EUSO
pixel and the corresponding single photoelectron pulse distribution are
shown in Figure~\ref{fig:S-curve}.  The black vertical line in the right
panel corresponds to the optimized threshold.

\begin{figure}[!ht]
	\centering
   \includegraphics[width=.49\textwidth]{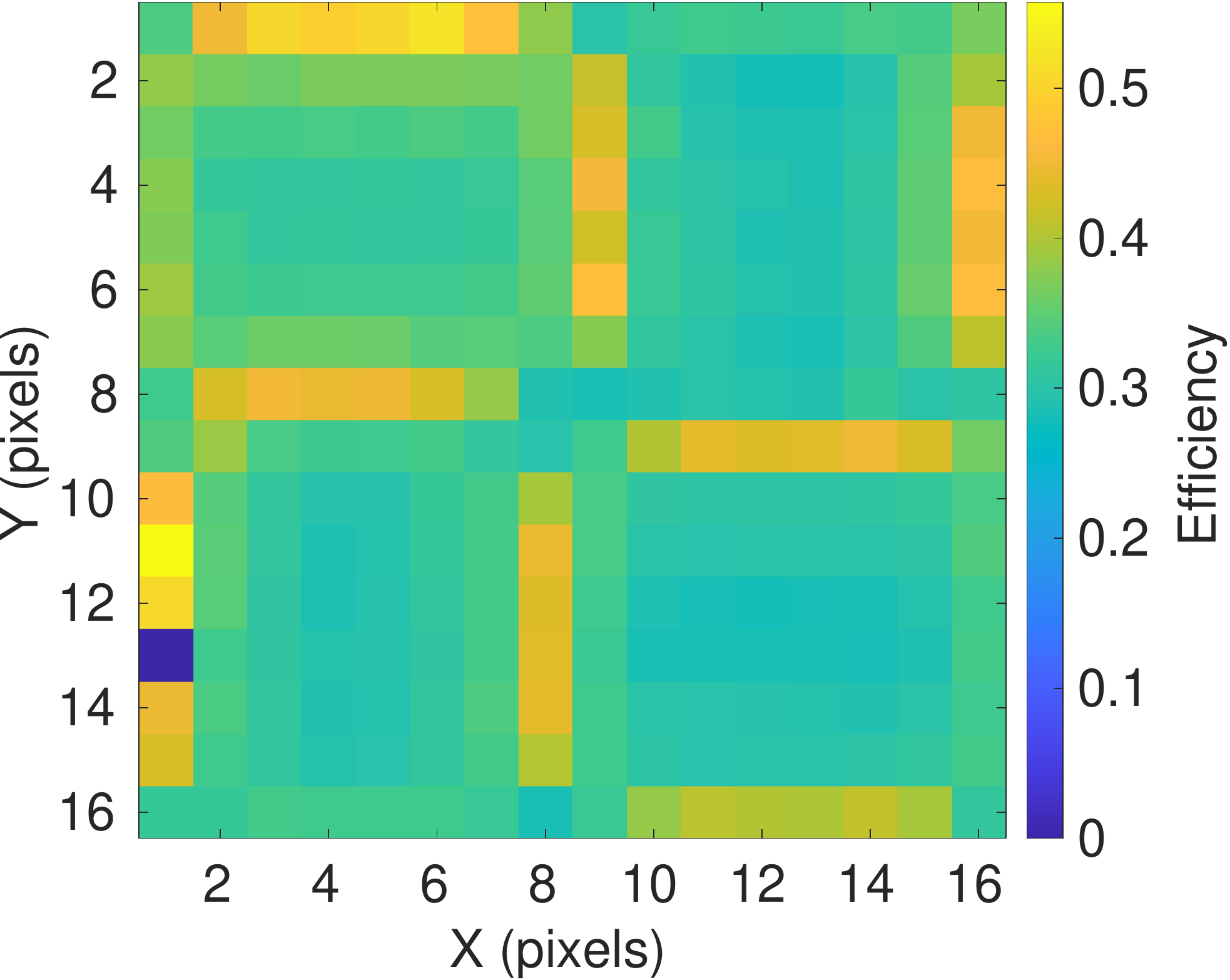}
   \caption{Efficiency map for one EC unit.
		Colors denote estimated photon detection efficiency of individual
		pixels in fractions of~1, which corresponds to 100\% efficiency.
		}
   \label{fig:PDM_eff}
\end{figure}

Measurements with various incoming UV intensities were performed to
obtain the dependence of the count rate on the number of photons.  At
the region of low intensities, this curve is linear, and then it begins
to decline due to the pile-up effect. The curve allows us to obtain
photon detection efficiency in a linear region, as well as an estimation
of a dead time of photon counting determined by a SPACIROC3 operation.
The map and distribution of channel efficiencies are shown in
Figure~\ref{fig:PDM_eff}. It can be seen that the efficiency of the
majority of the pixels lies in the range of 20--40\%. A number of border
pixels have a higher noise due to their position near the border of the
MAPMT. Therefore, their efficiency is overestimated, which can be seen
in the pixel map.  They have higher count rates and this should be taken
into account in trigger algorithms because this can cause false
triggers.  (Pixel (1,~13) in this PDM is not functional.) The average
value of the efficiency of the recording channels with preset individual
thresholds for the registration of single-photoelectron pulses for the
given EC-unit equals~34\%.

The CPU is placed inside the ISS. The CPU consists of two subsets: the
main one and the backup one. A cold standby scheme is provided, in which
only one CPU subset works at a time. Both subsets of the CPU are
implemented in one housing. Each CPU subset is designed on two printed
circuit boards: a power supply board and a processor board. 

The CPU power supply board contains a secondary power supply and a
microcontroller. The microcontroller implements algorithms for the
sequence of voltage supply to the processor board. In addition, the CPU
microcontroller is the master device on the CAN bus, through which,
commands are sent to turn on the components of the telescope and
telemetry information is collected. The overall dimensions of the CPU
are $400\times300\times90$~mm$^3$, which meets the requirements and
allows it to be freely transported through the hatches of the ISS and
the Progress cargo. It is equipped with two touch screens that allow
astronauts to monitor the status of all K-EUSO subsystems and operate a
manual control if necessary.

%______________________________________________________________________________
\section{Expected Performance}
\label{sec:performance}

In this section, estimations of the scientific performance of \keuso{}
are presented following~\cite{Fenu:2021wub}.  The study is based on
intensive simulations with the ESAF framework~\cite{Fenu:2019+0} and
their consequent analysis.
The studies presented here must be regarded as preliminary since the
configuration of the detector is still in the course of definition.
However, both the trigger and reconstruction performance should be
considered as indications of the observatory performance.

\begin{figure}[!ht]
	\centering
	\includegraphics[height=.3\textheight]{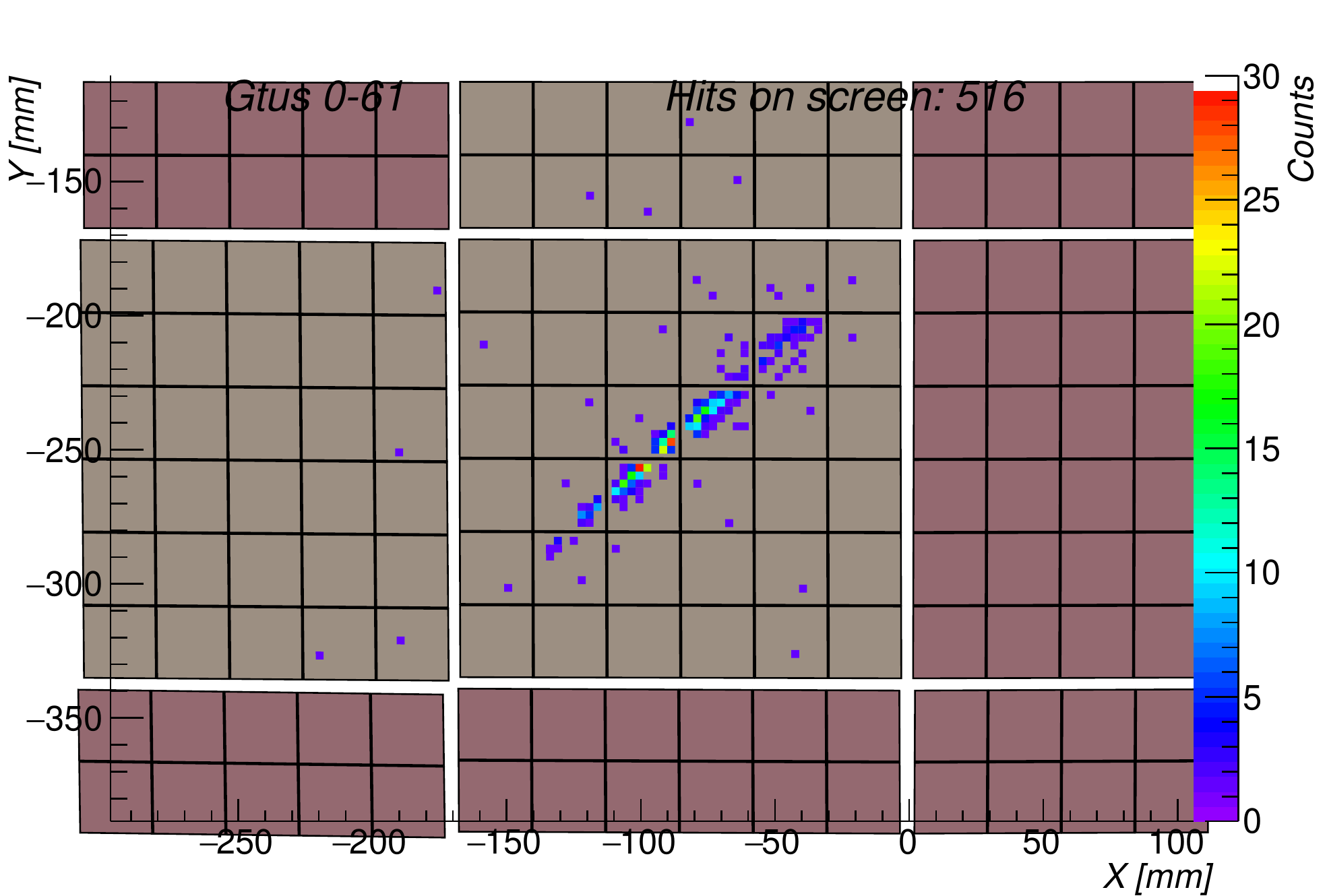}

	\includegraphics[height=.2\textheight]{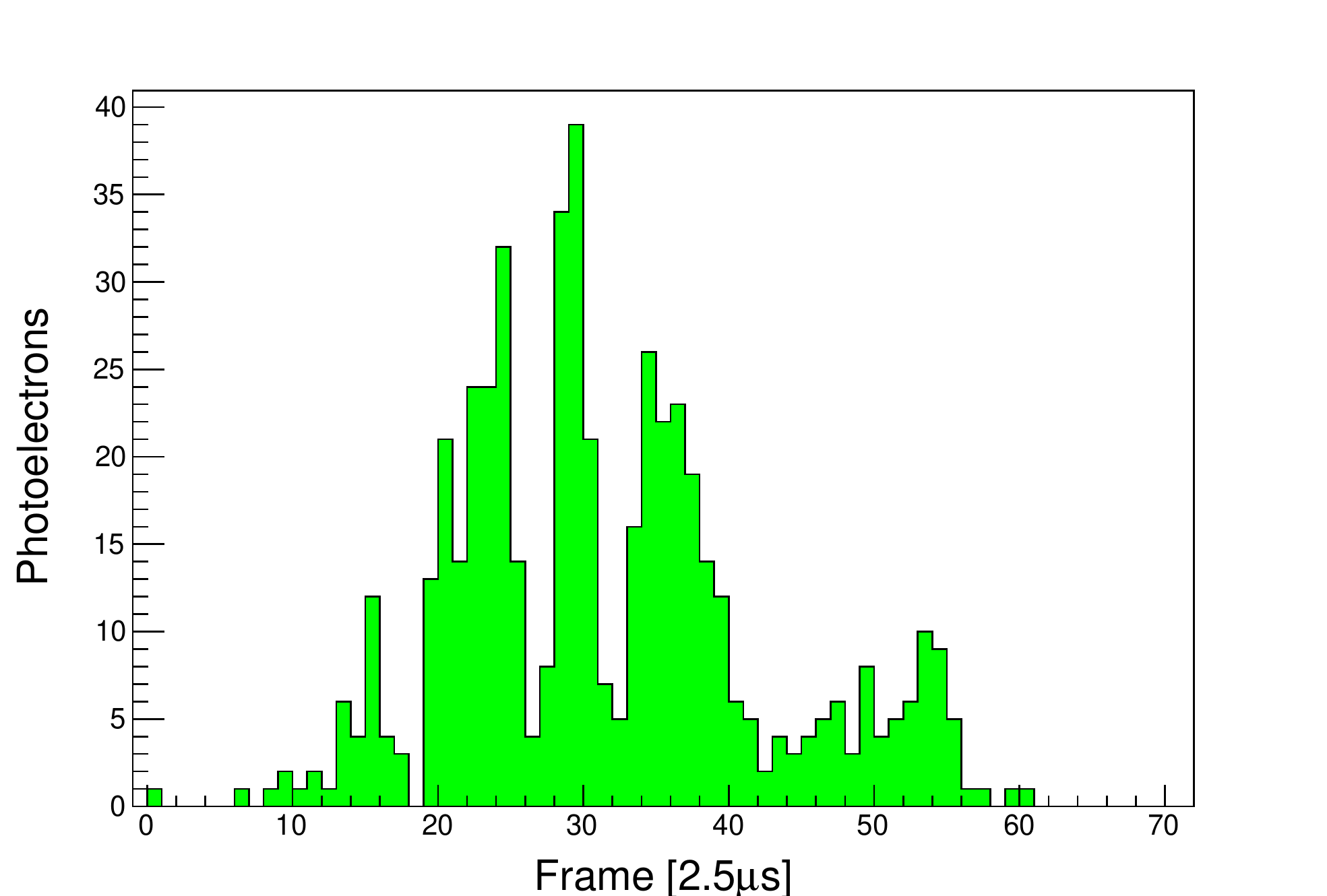}\quad
	\includegraphics[height=.2\textheight]{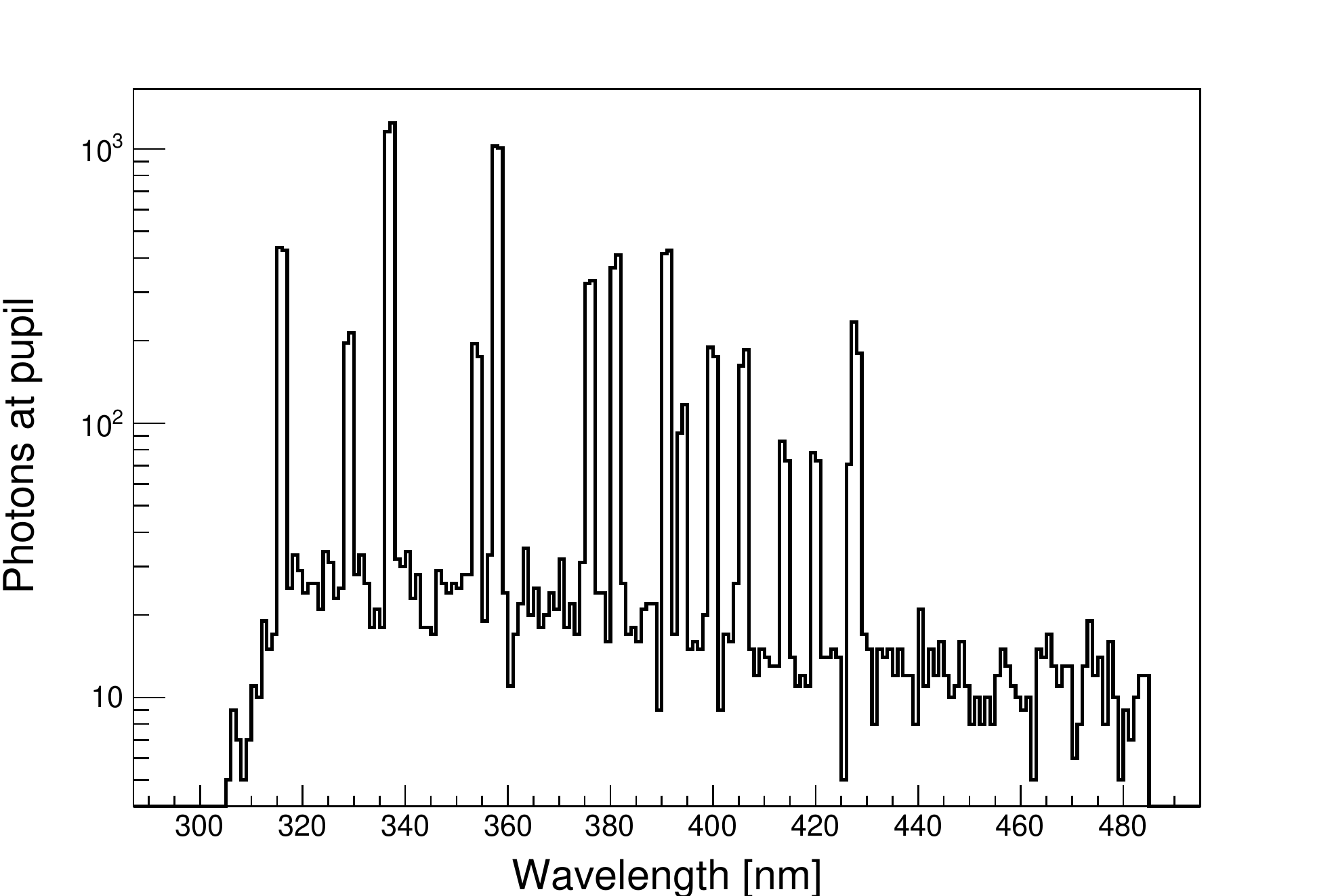}

	\caption{Distribution of photoelectrons from an EAS
		generated by a 100~EeV proton arriving at the zenith angle
		of $60^\circ$ simulated with ESAF. Only signal
		from the shower is shown.  Top: distribution of the signal
		on the focal surface.  Bottom left: the
		corresponding time distribution of photoelectrons.
		Bottom right: the spectrum of photons at the detector entrance.}

	\label{image:ESAF1e20_60}
\end{figure}

An example of the \keuso\ response to a \uhecr\ simulated with ESAF is
shown in Figure~\ref{image:ESAF1e20_60}. The top panel demonstrates a
distribution of photoelectrons from an EAS generated by a 100~EeV proton
arriving at the zenith angle of $60^\circ$ (without any airglow emission
taken into account) on the focal surface.  The same photoelectrons are
plotted as a function of time in the bottom left panel.  A periodic
decrease in the signal intensity is caused by the gaps between MAPMTs.
The bottom right panel shows the wavelength spectrum of photons entering
the detector.  The fluorescence emission lines can be seen together with
the continuum Cherenkov emission. 

%_______________________________________________________________________
\subsection{Exposure}

The trigger algorithms of the \keuso{} mission have been developed in
the framework of the JEM-EUSO program~\cite{2017NIMPA.866..150A} and are
currently in the process of optimization. {The results presented in
the following are obtained directly using the scheme adopted for
JEM-EUSO.}  The logic is structured in a number of stages, each reducing
the trigger rate by several orders of magnitude.  The first level
trigger, to be operated at the level of the PDM, looks for
concentrations of the signal localized in space and time.  {An
excess of signal is required in cells of $3\times3$ pixels integrated
over five consecutive GTUs. Thresholds are set to have a spurious trigger
rate, due to nightglow fluctuations, below 10 Hz/PDM.} The second level
trigger is activated each time the first level trigger conditions are
satisfied and integrates the signal intensity in a sequence of preset
directions that cover the entire phase space.  {Every GTU, the
detected counts are summed in blocks of $3\times3$ pixels along the
preset direction for a total of 15 consecutive GTUs around the location
and GTU where the first trigger level has been activated.} Whenever the
integrated signal along a direction overcomes a preset threshold, the
second level trigger is issued.  The activation of the second level
trigger starts the transmission and data storage procedure.  The data
acquisition is therefore stopped and data are either saved on a hard
disk or sent to ground by telemetry.  This way, the trigger reduces the
data flow by several orders of magnitude.  Thresholds {at the second
trigger level} are set to have a rate of the trigger every few seconds
at most {due to nightglow fluctuations} to make the data acquisition
consistent with the telemetry budget. {In the case of JEM-EUSO, these
trigger conditions were satisfying a telemetry budget of 300 kbit/s. The
lower number of \keuso\ PDMs would allow for a slightly higher trigger rate,
assuming similar telemetry. The optimization of the trigger parameters
(thresholds and number of integrated GTUs) will be carried out in future, when
the project proceeds to the next development phases.} The aim
of this section is therefore to test the efficiency curve of the
algorithm with respect to cosmic ray air showers.

The exposure calculation is based on a Monte Carlo simulation of the EASs of
variable energy and direction.  To avoid border effects, cosmic rays are
injected in an area~$A_\mathrm{simu}$ larger than the FOV of the
detector.  The ratio of the triggered $N_\mathrm{trigg}$ over simulated
events $N_\mathrm{simu}$ is then calculated for each energy bin. The
solid angle~$\Omega$ from which cosmic rays arrive on the field of view
is also included in the formula.  The effects of the day--night cycle and
moon phases are taken into account in~$\eta$, the astronomical duty
cycle.  The effects of clouds and artificial lights are taken into account
by $\eta_\mathrm{clouds}$ and $\eta_\mathrm{city}$, respectively.  In
this formula, we assumed $\eta$ = 0.2, $\eta_\mathrm{clouds} = 0.72$ and
$\eta_\mathrm{city}$ = 0.9, as estimated in~\cite{JEM-EUSO-exposure}. The
exposure~$\mathcal{E} (E)$  is then calculated over time~$t$, which is
assumed to be~1 year in the following:

\begin{equation} \label{eq:exposure}
  \mathcal{E} (E) = \frac{N_\mathrm{trigg}}{N_\mathrm{simu}} (E)
		\times A_\mathrm{simu} \times \Omega \times \eta \times
		\eta_\mathrm{clouds} \times \eta_\mathrm{city} \times t.
\end{equation}

\begin{figure}[!ht]
	\centering
	\includegraphics[height=.3\textheight]{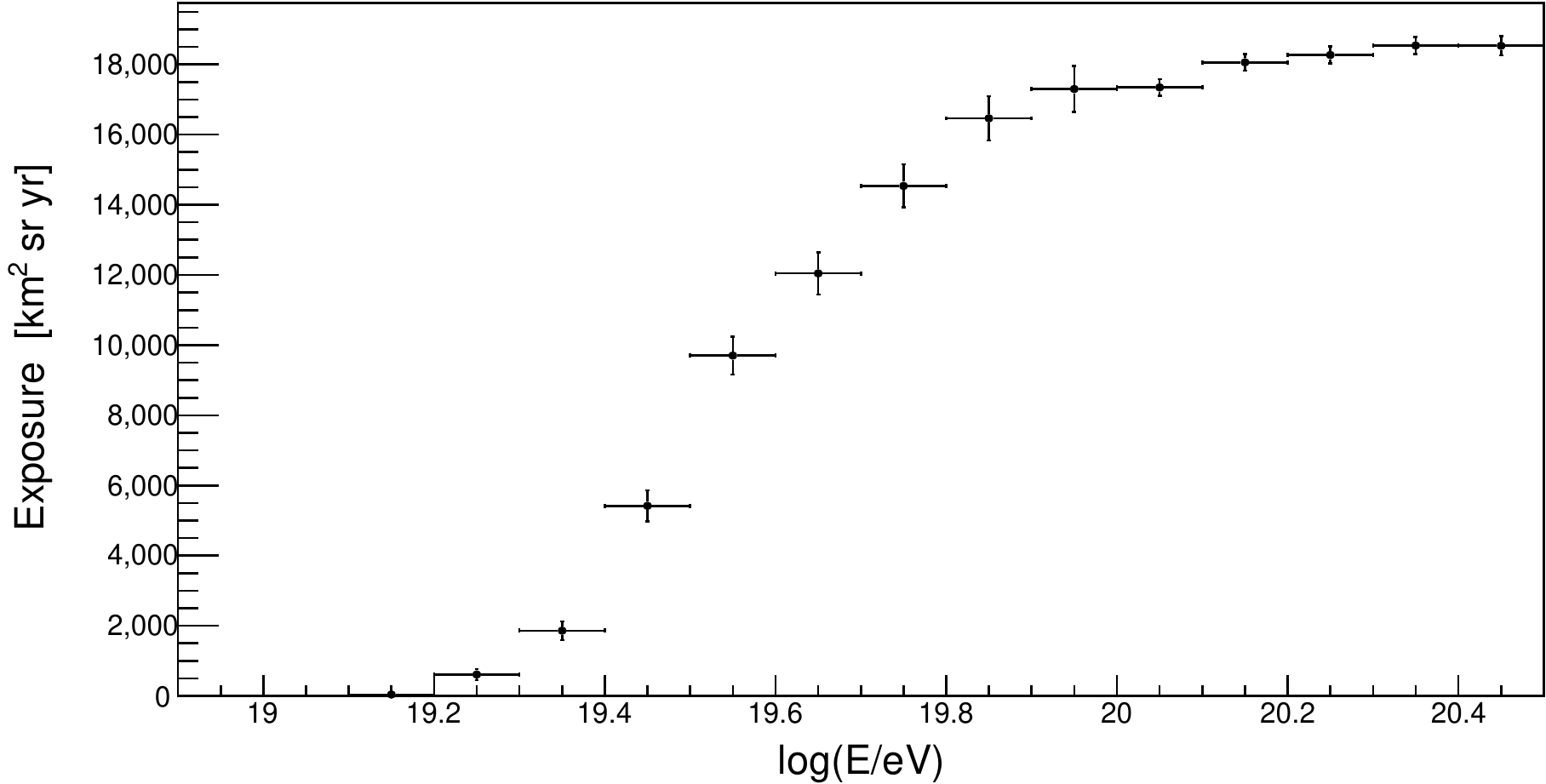}

	\caption{Annual exposure of \keuso{} as a function of UHECR primary
	energy.}

	\label{fig:exposure}
\end{figure}

The yearly exposure as a function of energy is shown in
Figure~\ref{fig:exposure}.  It can be seen that \keuso{} achieves an
exposure of $\sim$18{,}000~km$^2$~sr per year at the plateau, which is
reached at energies above 100~EeV {(to be compared with
5000--7000~km$^2$~sr per year of the Auger collaboration and the
TA$\times$4 of Telescope Array)}.  The 50\% efficiency is reached around
40~EeV.  Assuming the Auger spectrum~\cite{PierreAuger:2020qqz}, the
expected rate of \uhecrs\ triggered by \keuso\ is estimated to be around
65 events/year above 50~EeV, including~4 events with energies above
100~EeV.  For comparison, the Pierre Auger collaboration has
detected, on average, $\sim$19 events per year above 50 EeV for the SD
spectrum under 60 degrees.

%_______________________________________________________________________
\subsection{Angular Reconstruction}

At the occurrence of a trigger, the acquisition is stopped and data are
retrieved.  The information collected at this point is then used to
reconstruct parameters of the primary~particle.

The first step consists of the recognition of the track, namely, of
pixels and frames in which the light of the shower arrives.  A
comprehensive review of the signal identification methods is given
in~\cite{EA_Angle,EA_Energy}.  All of the methods look for
concentrations of the signal in space and time that exhibit kinematics
consistent with the presence of an extensive air shower.

\begin{figure}[!ht]
	\centering
	\includegraphics[width=.48\textwidth]{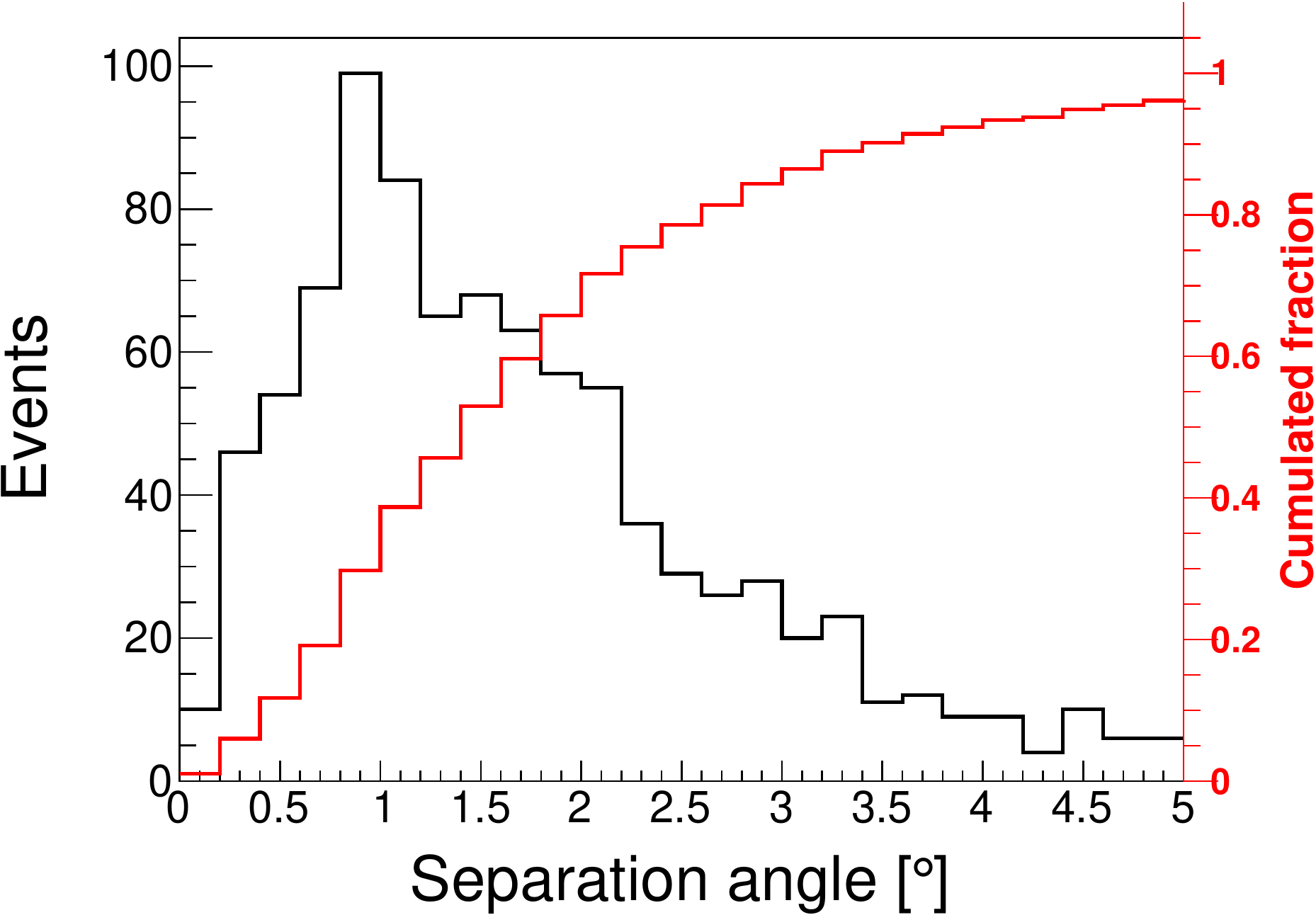}\quad
	\includegraphics[width=.48\textwidth]{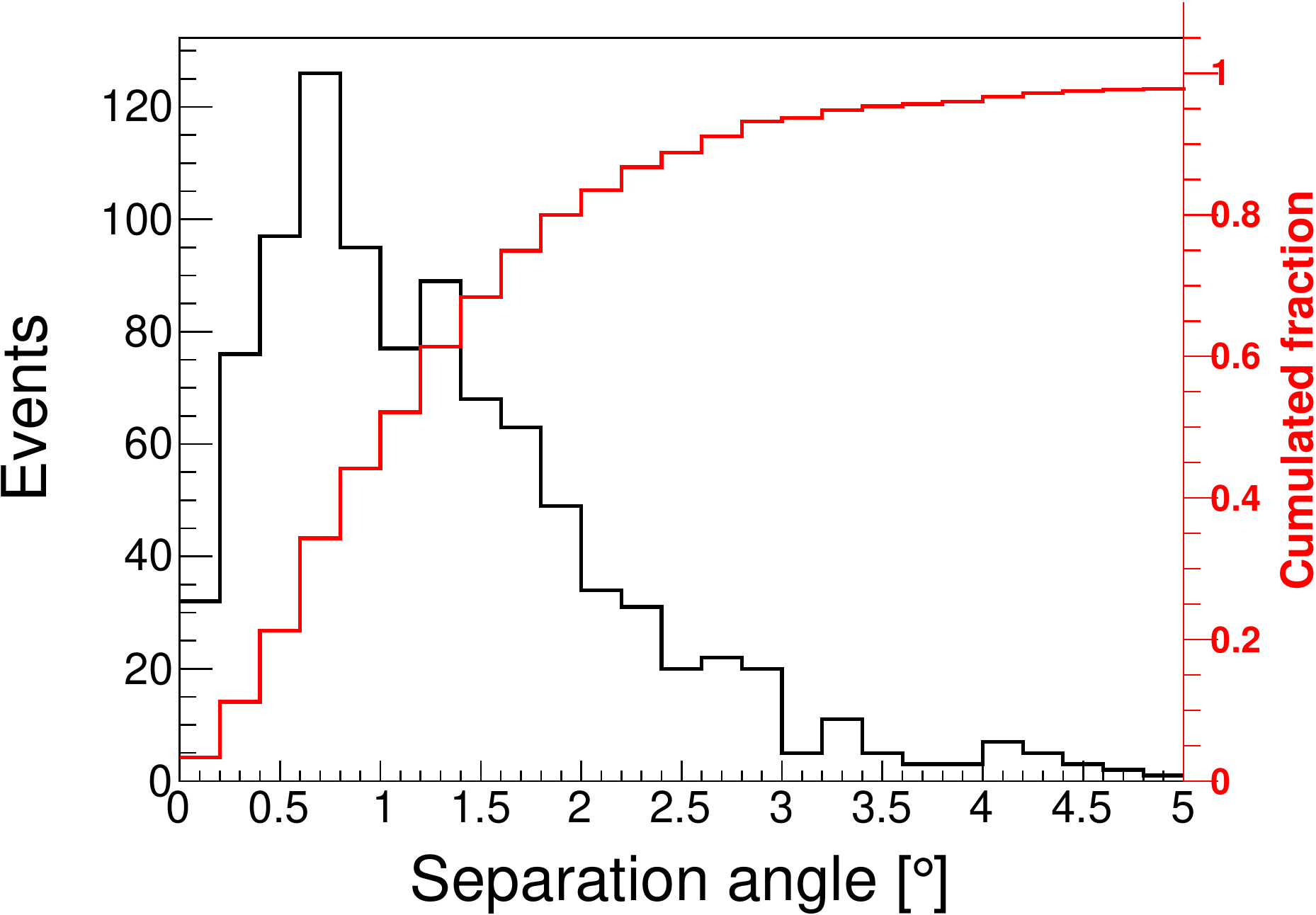}

	\caption{Performance of the angular reconstruction for UHECRs with
	the energy 100~EeV in the center of the field of view. Left:
	45$^\circ$ zenith angle.  Right: 60$^\circ$ zenith angle.}

	\label{fig:AngleReco}
\end{figure}

The angular reconstruction extracts the arrival direction of the primary
particle from the distribution and timing of the identified track.
Several methods have been tested in the context of the JEM-EUSO
program~\cite{EA_Angle}.  The method used for this work is based on a
$\chi^2$ fit of the position and timing of the shower signal (the
so-called \textit{Numerical Exact~1} method).  In this method, the
identification of the plane where both the shower and the detector lie
is the first step of the procedure.  The zenith angle of the shower is
then reconstructed by comparing the arrival time of photons from a test
shower and the identified track.  The reconstruction performance for the
arrival directions of EAS generated by 100~EeV protons arriving at
zenith angles of~$45^\circ$ and $60^\circ$ in the center of the field of
view are shown in Figure~\ref{fig:AngleReco}.  To assess the quality of
the reconstruction, we plot the integral of the event distribution
from~0 to a specific angle (in red). Here, 68\% of the events fall
within $1.5^\circ\text{--}2^\circ$, proving an excellent reconstruction
performance at the highest energies.

\begin{figure}[!ht]
	\centering
	\includegraphics[height=7cm]{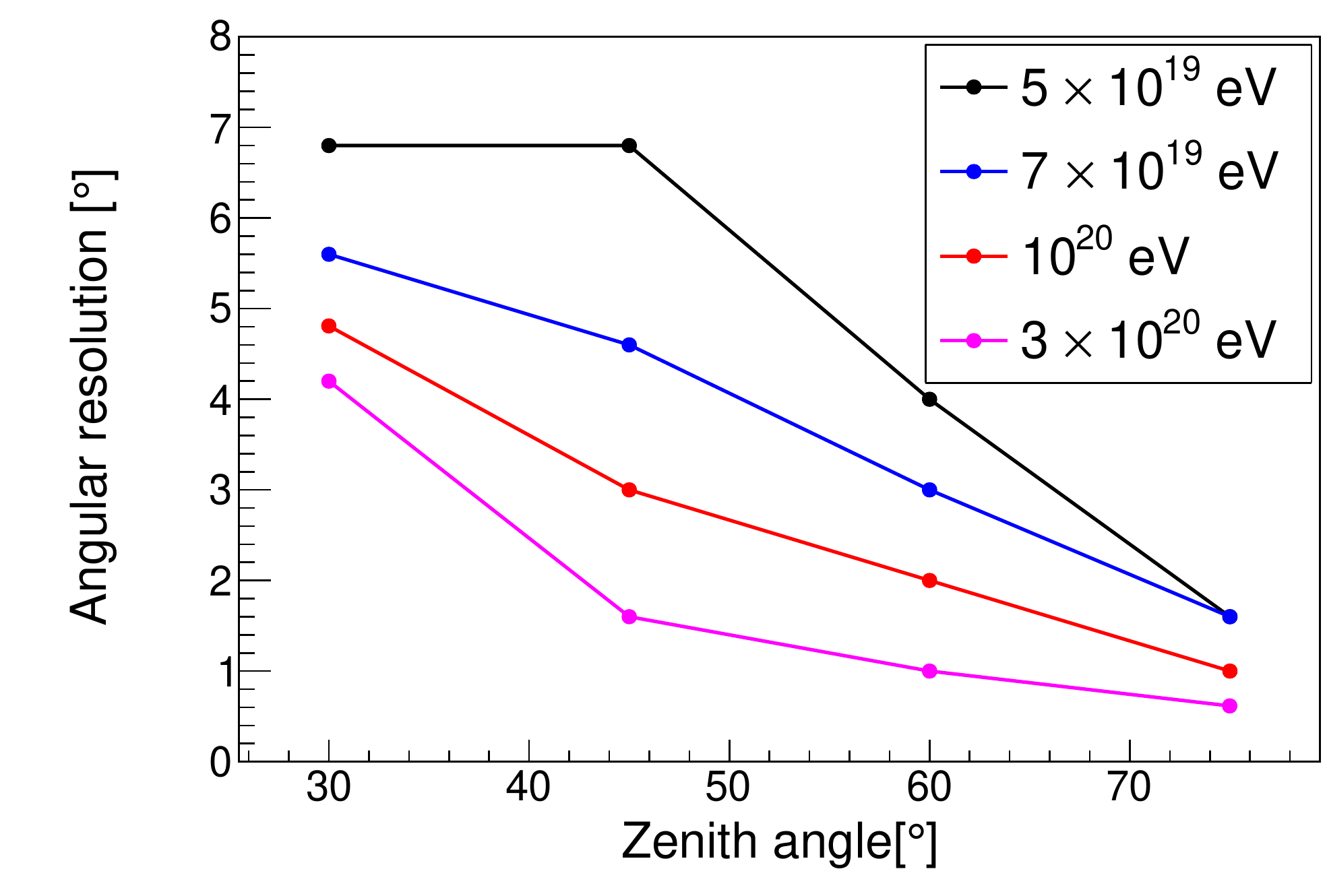}

	\caption{Estimated angular resolution of \keuso{} for different
	zenith angles and energies of a primary particle
	in the full field of view of the detector.}

	\label{fig:AngleRes_ALL_FULL}
\end{figure}

In Figure~\ref{fig:AngleRes_ALL_FULL}, the angular resolution is plotted
as the angle within which 68\% of events fall.  For this plot, we
simulated 500 EASs in 16 different combinations of the energy and zenith
angle.  For each condition, the events were simulated over the entire
field of view of the detector.  It can be seen that \keuso{} will have a
resolution between~$4^\circ$ to~$7^\circ$ for small zenith angles, but
it improves to $1^\circ\text{--}2^\circ$ for nearly horizontal events.
There is a clear improvement trend as the energy increases. For
comparison, the Auger collaboration achieves an angular reconstruction
better than 1 degree, whereas TA achieves a resolution of the same
order.

%_______________________________________________________________________
\subsection{Energy Reconstruction}

The energy reconstruction is performed according to~\cite{EA_Energy} and
is based on the signal identified by pattern recognition.  The
photoelectron profile is reconstructed based on the counts falling in
the identified track with the airglow subtracted on average.  The
attenuation occurring in the detector is corrected following a look-up
table relating an incident direction and wavelength of photons to the
detector efficiency.  Several methods to reconstruct the shower geometry
have been implemented and are discussed in~\cite{EA_Angle}.  With a
determination of the position of the shower in the atmosphere, it is
possible to calculate the amount of atmospheric extinction and the
luminosity curve.  Knowledge of the fluorescence yield is then used to
reconstruct the charged particle profile of an EAS.  Such a profile is
then fit with a shower profile parameterization to obtain the energy and
the depth of the maximum\footnote{The function used here is the so
called Gaisser--Ilina--Linsley (GIL) function. Simulations are being
updated to include more up to date parameterizations and MC shower
simulators as well.}.

\begin{figure}[ht!]
	\centering
	\includegraphics[width=.48\textwidth]{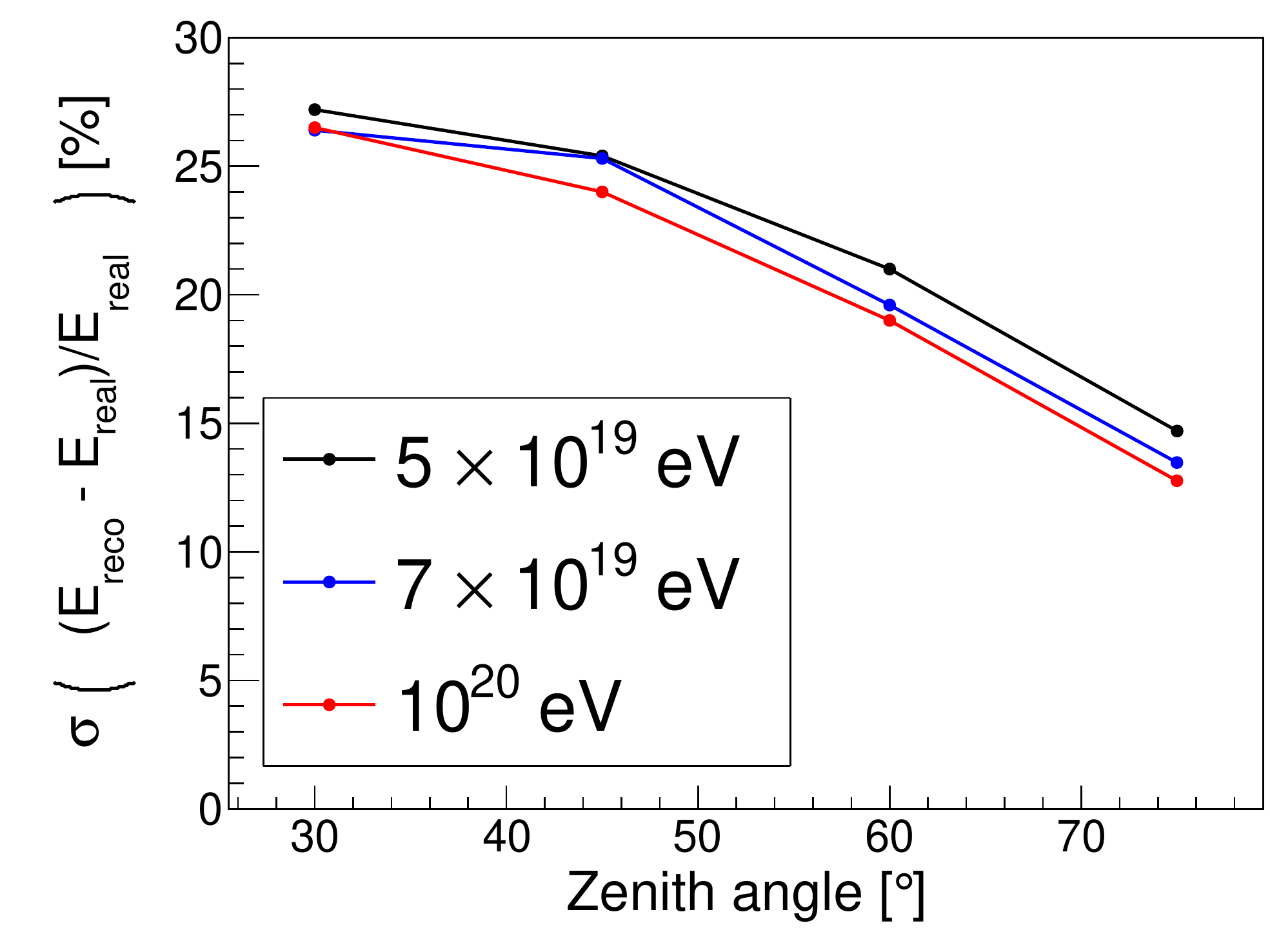}\quad
	\includegraphics[width=.48\textwidth]{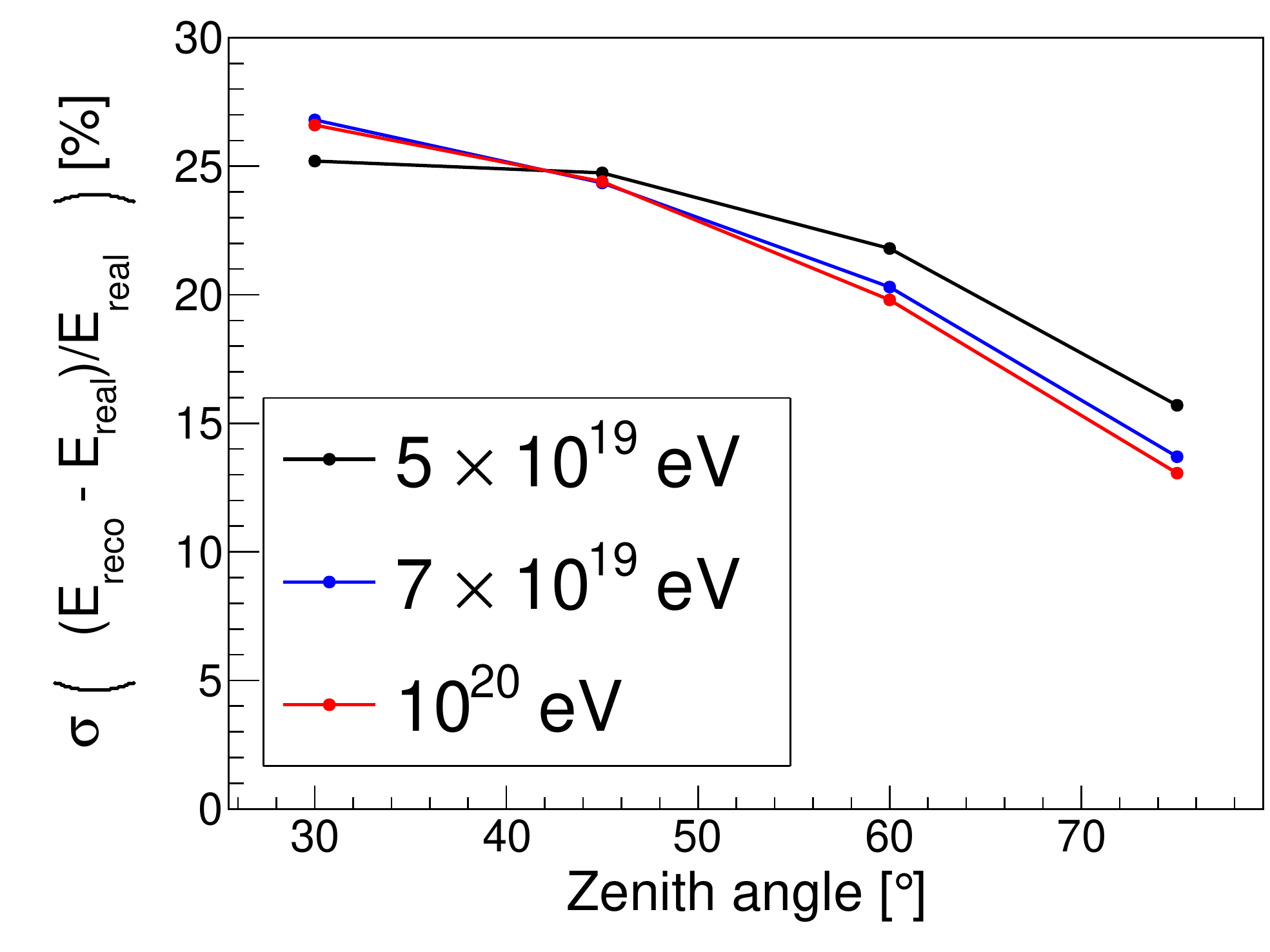}

	\caption{Estimates of energy reconstruction for different energies and
	zenith angles. Left: center of the field of view. Right: full field of
	view of the detector. }

	\label{fig:EnergyRes}
\end{figure}

\begin{figure}[!ht]

	\centering
	\includegraphics[width=.49\textwidth]{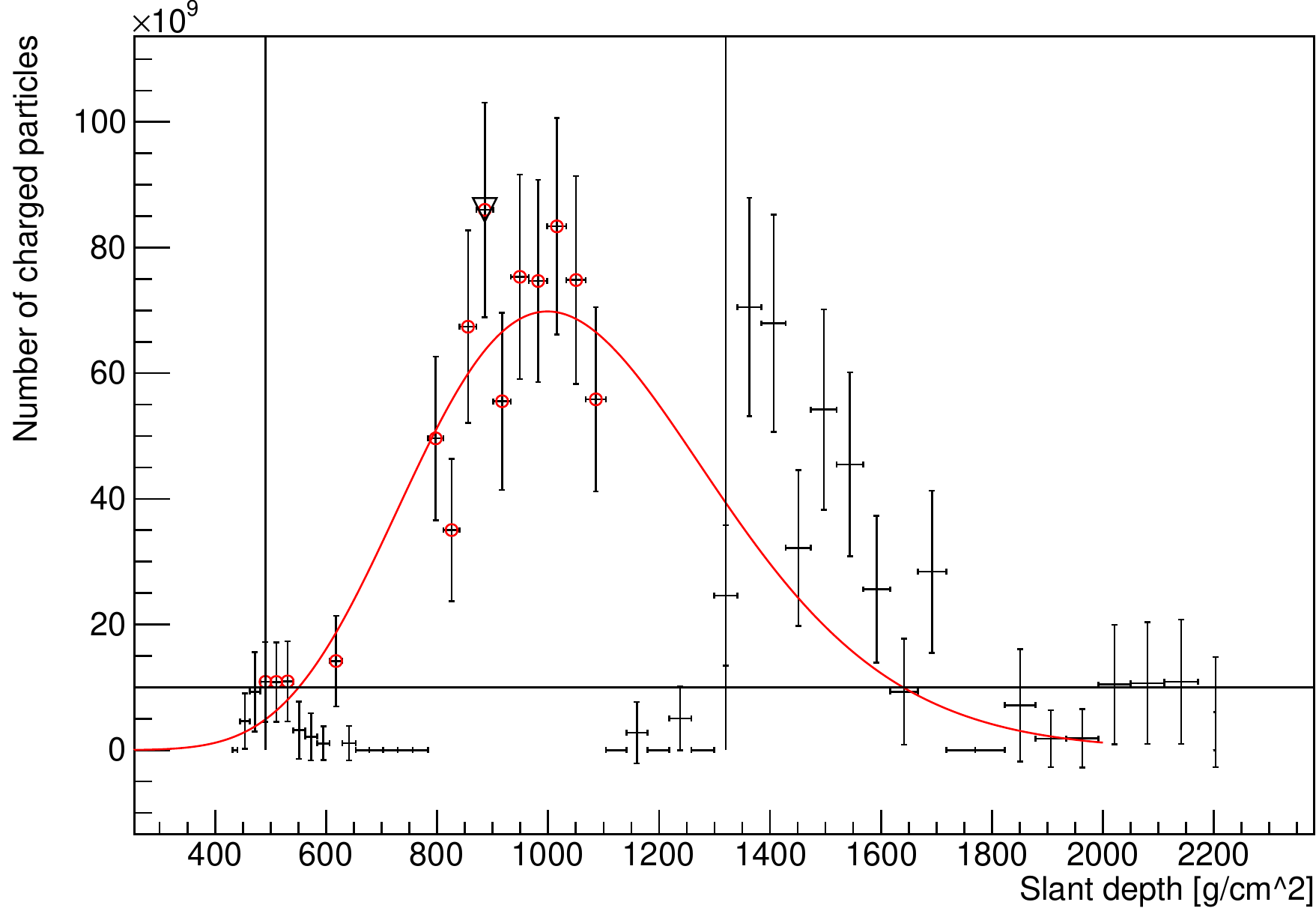}
	\includegraphics[width=.49\textwidth]{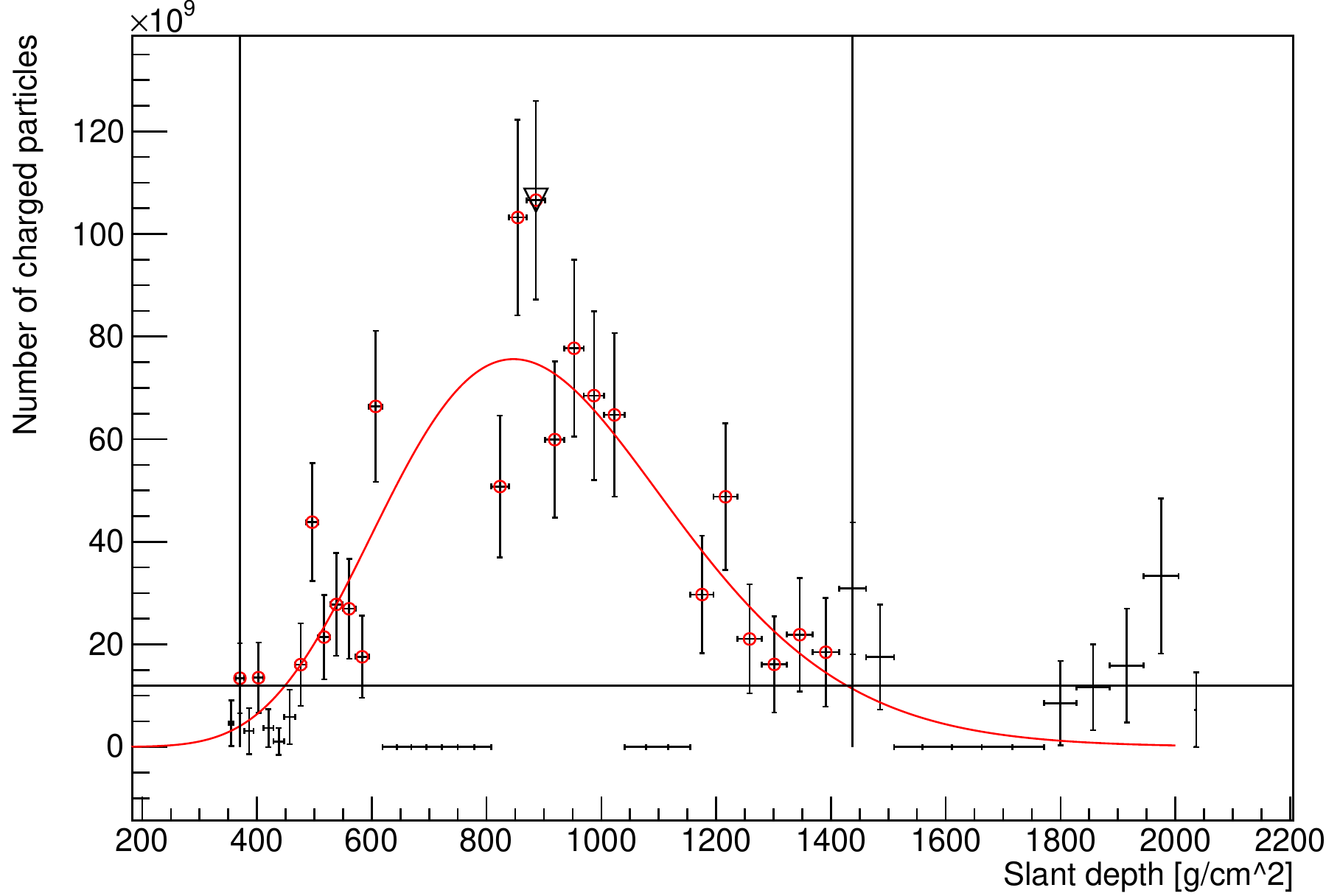}
	
	\caption{Examples of reconstructed shower profiles as a function
	of slant depth.  Both showers were generated
	by simulated 100~EeV \uhecrs\ arriving at $60^\circ$ zenith angle.
	The red curves indicate shower profile fits.
	Red circles mark data points used for the reconstruction.
	Vertical black lines show boundaries of the interval selected for
	the reconstruction by the algorithm. Horizontal black segments
	correspond to gaps between PDMs and/or PMTs.}
	
	\label{fig:Profile}

\end{figure}

Estimations of the energy resolution of \keuso{} for UHECRs with
different energies arriving at various zenith angles are shown in
Figure~\ref{fig:EnergyRes}. The results were obtained with 2500 showers
simulated at fixed energies and zenith angles, both for the center and
for the full field of view of the detector.  The resolution was
estimated as the standard deviation of the $\delta E = (E_\mathrm{reco}
- E_\mathrm{real}) / E_\mathrm{real}$ distribution {which can be well
approximated by a Gaussian}\footnote{A residual bias of $\sim$10\% is
remaining in the distribution. The bias is still under investigation and
will be corrected in the future.  }.  It can be seen that the energy
resolution is around 25\% at low zenith angles and improves up to around
15\% for nearly horizontal events, with a small improvement trend
towards higher energies.  No significant decrease in the performance was
observed for events simulated in the full field of view. For comparison,
Auger has an energy resolution above $10^{19}$ eV of the order of 7$\%$,
whereas TA has 18$\%$ in the same energy range.  The systematic
uncertainties on the energy scale are 14\% and 21$\%$ for Auger and TA,
respectively.  Two examples of reconstructed profiles of the signal from
100~EeV \uhecrs\ arriving at the zenith angle of $60^\circ$ are shown in
Figure~\ref{fig:Profile} as black crosses.  Fits of the shower
parameterization are shown in red.  The two showers are different in the
location of the signal in the focal surface.  The one shown in the left
panel crosses fewer gaps between MAPMTs than that in the right panel,
and the maximum of the profile has enough data points for an accurate
reconstruction.  As a result, we found $\delta E=0.01$ for the shower in
the left panel and $\delta E = 0.1$ for the one in the right panel.

A totally different approach to the recognition of patterns produced by
UV emission of EASs on the focal surface and to the reconstruction of
parameters of primary \uhecrs\ can be based on machine learning methods.
A high efficiency of neural networks for identifying certain types of
signals in the TUS data was demonstrated recently~\cite{BMY2020,
2021Univ....7..221Z}. The first applications of neural networks to CR
parameter reconstruction are also potentially interesting; see, e.g.,
\cite{2018APh....97...46E, Ivanov:2020nfo, PierreAuger:2021fkf}.  A
promising method of identifying EAS candidates in the data with the help
of neural networks in real time has been developed for the EUSO-SPB2
mission~\cite{Filippatos:2021b5}.  The method will allow one to
effectively suppress a big number of false positives, which is important
in conditions of a limited telemetry budget.  The procedure will be
thoroughly tested during the flight that is planned to be performed in
2023~\cite{Eser:2021mbp}.  This approach will be examined during the
future stages of development of the \keuso\ mission.

\subsection{Depth of Maximum Reconstruction}

A reconstruction of the depth of the maximum of an EAS~$X_\mathrm{max}$ is
also performed according to~\cite{EA_Energy} and is obtained from the fit of
the reconstructed shower profile.  A shower maximum can clearly be
identified from the profiles shown in Figure~\ref{fig:Profile}.  In this
work, we only present a few examples of the reconstruction performance
obtained in some specific condition.  The method we use here considers
only events where a Cherenkov reflection peak is visible. In this way,
the impact point of the shower can be identified in the profile, and
therefore a clear constraint can be put onto the shower geometry.  For
all of the other events, a more complex iterative study is needed and
further work will be published in the future.

\begin{figure}[ht!]
	\centering
	\includegraphics[width=.48\textwidth]{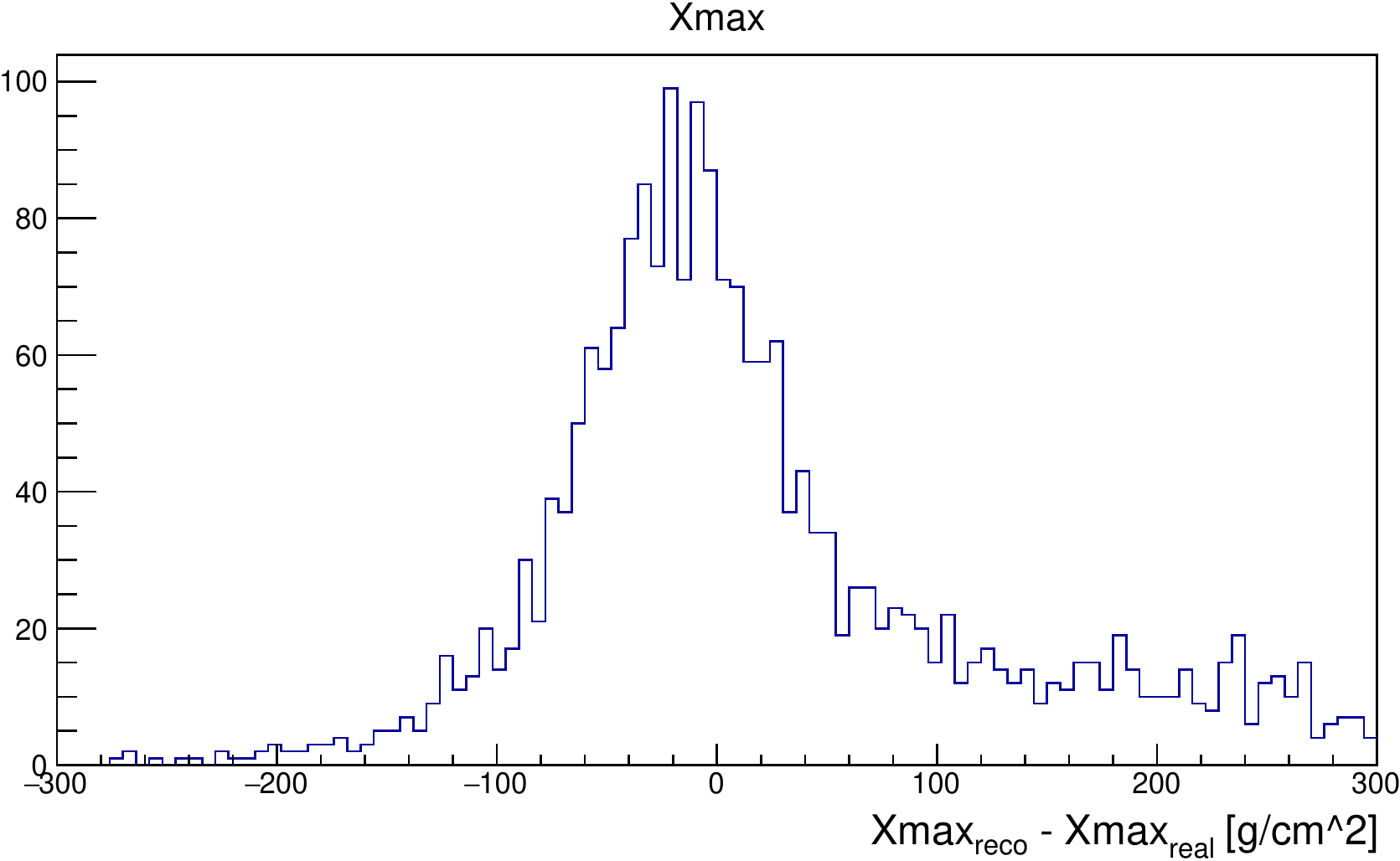}\quad
	\includegraphics[width=.48\textwidth]{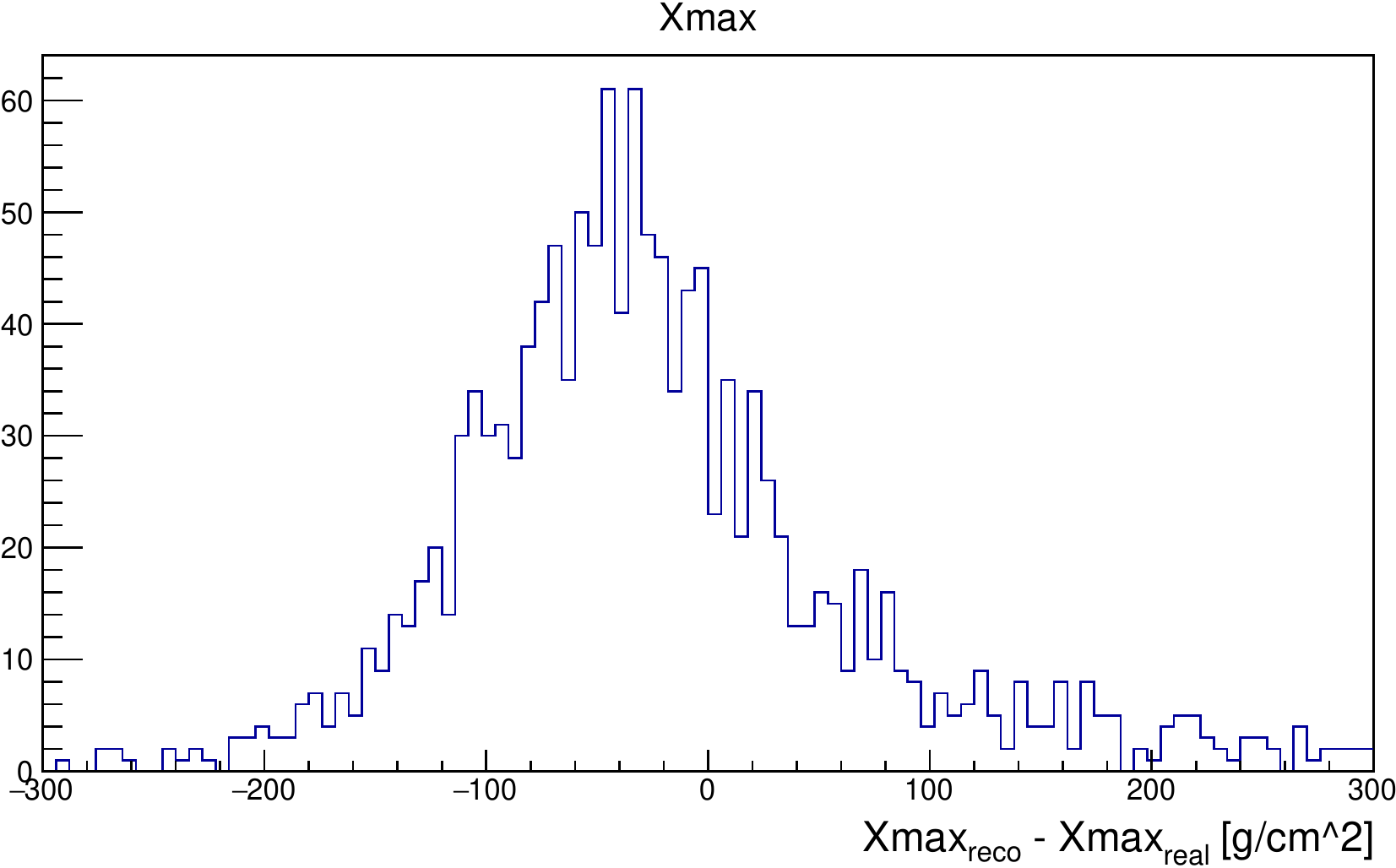}

	\caption{Estimates of $X_\mathrm{max}$ reconstruction for 200~EeV.
	Left: $30^\circ$ zenith angle. Right: $45^\circ$ zenith angle.}

	\label{XmaxExamples}
\end{figure}

The $X_\mathrm{max}$ resolution expressed as
$X_\mathrm{max}^\mathrm{reco} - X_\mathrm{max}^\mathrm{real}$ in units
of g/cm$^2$ is shown in Figure~\ref{XmaxExamples} for 200~EeV and for
$30^\circ$ and $45^\circ$ in the left and right panels, respectively.
The width of the distributions is around $\pm60$~g/cm$^2$, whereas the
mode of the distribution varies from approximately $-20~\text{g/cm}^2$
to $-40~\text{g/cm}^2$.  The reason for this systematic bias is being
investigated and will be corrected once the detector configuration is
defined completely.  The long tails on the right side of the
distributions are due to events impacting on the ground before the
maximum is reached. The left tails are, on the other hand, due to events
where the maximum is lost because of inter-PMT gaps. The former case is
indeed more pronounced for vertical events, for which, the shower
reaches the ground earlier in the shower development.  A residual bias
of the distribution is still visible and will be addressed in future
publications.  An overview of the performance in certain conditions is
given in Table~\ref{parametersReco}. The resolution is always around
50--90~g/cm$^2$ for the center, and similar values are obtained for the
whole field of view. 

\begin{table}[!ht]
    \begin{center}
    \caption{$X_\mathrm{max}$ resolution of the K-EUSO detector in the center
		 of the field of view.}
    \smallskip
    \begin{tabular}{|c|c|c|}
    \hline
        Energy [EeV] & Zenith angle [$^\circ$] & Resolution [g/cm$^2$] \\
    \hline
        70 & 30 & 83 \\
    \hline
        70 & 45 & 94 \\
    \hline
        100 & 30 & 69 \\
    \hline
        100 & 45 & 75 \\
    \hline
        200 & 30 &  50 \\
    \hline
        200 & 45 &  66 \\
    \hline
    \end{tabular}
    \label{parametersReco}
    \end{center}
\end{table}

These results should be regarded as preliminary, given the challenging
nature of the $X_\mathrm{max}$ reconstruction in monocular mode. Future
studies will specifically address the $X_\mathrm{max}$ reconstruction
performance. For comparison, Auger achieves 15~g/cm$^2$ above
$10^{19}$~eV for the resolution and 10~g/cm$^2$ for the systematic
uncertainty. TA achieves, very similarly, 17~g/cm$^2$ for the resolution
and 17~g/cm$^2$ for the systematic uncertainties.

%______________________________________________________________________________
\section{Discussion and Conclusions}

\keuso\ is an orbital telescope for studying \uhecrs, and is planned to
be deployed on the Russian segment of the International Space Station in
the near future for a 2-year mission (or a longer mission if operations
of the ISS are extended to 2030).  With its latest design presented
above, the instrument fits nicely in the Progress cargo and can be
comparatively easily deployed outside the ISS.  A number of the key
elements of \keuso\ are already in production.

Despite the strongly reduced optical system, \keuso\ possesses technical
parameters that are sufficient to make an important step in revealing
the nature and origin of \uhecrs.  With the expected rate of triggered
events of around 65 per year for energies above 50~EeV (assuming the
Pierre Auger Observatory spectrum) and the energy resolution varying
from~15\% to~25\% for different zenith angles, \keuso\ will allow us to
verify certain results of Auger and the Telescope Array concerning the
energy spectrum of cosmic rays at respective energies.  With a
sufficiently long mission, \keuso\ will provide an opportunity to test
the dependence of the energy spectrum on the declination reported by the
TA.

With its full-sky exposure, it will provide the first complete map of
the sky in \uhecrs\ obtained in a consistent way by one single
instrument.  The angular resolution of the telescope will be good enough
to explore anisotropies at medium and large scales. Combined with high
statistics of events, this will provide a chance to verify the existence
of the hot spot observed by TA and the warm spot found by Auger, to look
for a possible correlation between the arrival directions of \uhecrs\
and the distribution of matter in the local Universe and to test if
there are signatures of a nearby source contributing to the flux of CRs
above $\sim$50~EeV in their large-scale anisotropy.  Despite the large
uncertainty on the depth of the shower maximum on individual events, the
uncertainty on the average composition will be reduced by the statistics
collected in each energy bin. Two energy bins could be envisaged: one
around the threshold (i.e., $5\times10^{19}$~eV) and the other at the
highest energies (i.e., close to $10^{20}$~eV). Both would contain
decent statistics and would be separated enough to account for effects
related to the energy resolution~\cite{Brummel:2013urn}.  Thus, the data
will also provide some information about the mass composition of cosmic
rays, which would allow for directly comparable measurements over the
entire sky for the first time.

Being a multi-purpose instrument with a huge field of view, unique
sensitivity and high temporal resolution, \keuso\ will be able to make
interesting contributions to atmospheric sciences, especially in regard
to transient luminous events, to meteor studies and to the search of
hypothetical nuclearites, as was demonstrated by comprehensive
simulations performed during the development of the JEM-EUSO
telescope~\cite{jem-euso-atmos,jemeuso-meteors} and confirmed later with
observations of TUS~\cite{2019RemS...11.2449K, nuclearites-ICRC2019,
Klimov:2021rxn} and Mini-EUSO~\cite{Mini-EUSO2021, Marcelli:2021uX,
Golzio:2021tv}.

%______________________________________________________________________________
\section*{Acknowledgments}

{\sloppy
The program of studying \uhecrs\ with orbital
telescopes was pioneered in Russia by Boris~Khrenov and
Mikhail~Panasyuk. Unfortunately, they passed away recently.
Their impact on the development of the research program was
profound and their scientific legacy is fully recognized by all their
collaborators.

}

The authors express their deep and collegial thanks to the entire
JEM-EUSO program and all its individual members.  The article was
prepared based on research materials carried out in the space experiment
``KLYPVE,'' included in the Long-term program of Experiments on board
the Russian Segment of the ISS.  This work was supported by the State
Space Corporation ROSCOSMOS, by the Italian Ministry of Foreign Affairs
and International Cooperation as a Project of Great Relevance between
Italy and Japan, by JSPS KAKENHI Grant (JP19H01915) and by the French
Space Agency, CNES.  The Russian authors are supported by the
Interdisciplinary Scientific and Educational School of Lomonosov Moscow
State University ``Fundamental and Applied Space Research.''

%______________________________________________________________________________
\bibliographystyle{JHEP}
\bibliography{keuso4arxiv}

\providecommand{\href}[2]{#2}\begingroup\raggedright\begin{thebibliography}{10}

\bibitem{Linsley1961}
J.~{Linsley}, L.~{Scarsi} and B.~{Rossi}, \emph{Extremely energetic cosmic-ray
  event}, \href{https://doi.org/10.1103/PhysRevLett.6.485}{\emph{Physical
  Review Letters} {\bfseries 6} (1961) 485}.

\bibitem{PierreAuger:2015eyc}
{\scshape Pierre Auger} collaboration, \emph{The {Pierre Auger} cosmic ray
  {Observatory}}, \href{https://doi.org/10.1016/j.nima.2015.06.058}{\emph{Nucl.
  Instrum. Meth. A} {\bfseries 798} (2015) 172}
  [\href{https://arxiv.org/abs/arXiv:1502.01323}{{\ttfamily
  arXiv:1502.01323}}].

\bibitem{TAx4-2021}
{\scshape Telescope Array} collaboration, \emph{Current status and prospects of
  surface detector of the {TAx4} experiment},  in \emph{Proceedings of 37th
  International Cosmic Ray Conference {\textemdash} PoS(ICRC2021)}, vol.~395,
  p.~203, 2021, \href{https://doi.org/10.22323/1.395.0203}{DOI}.

\bibitem{Deligny:2020gzq}
{\scshape Pierre Auger, Telescope Array} collaboration, \emph{The energy
  spectrum of ultra-high energy cosmic rays measured at the {Pierre Auger
  Observatory} and at the {Telescope Array}},  in \emph{Proceedings of 36th
  International Cosmic Ray Conference {\textemdash} PoS(ICRC2019)}, vol.~258,
  p.~234, 2020, \href{https://doi.org/10.22323/1.358.0234}{DOI}
  [\href{https://arxiv.org/abs/arXiv:2001.08811}{{\ttfamily
  arXiv:2001.08811}}].

\bibitem{Abbasi:2021qO}
{\scshape Pierre Auger, Telescope Array} collaboration, \emph{Joint analysis of
  the energy spectrum of ultra-high-energy cosmic rays as measured at the
  {Pierre Auger Observatory} and the {Telescope Array}},  in \emph{Proceedings
  of 37th International Cosmic Ray Conference {\textemdash} PoS(ICRC2021)},
  vol.~395, p.~337, 2021, \href{https://doi.org/10.22323/1.395.0337}{DOI}.

\bibitem{PierreAuger:2020qqz}
{\scshape Pierre Auger} collaboration, \emph{Measurement of the cosmic-ray
  energy spectrum above $2.5{\times} 10^{18}$~{eV} using the {Pierre Auger
  Observatory}}, \href{https://doi.org/10.1103/PhysRevD.102.062005}{\emph{Phys.
  Rev. D} {\bfseries 102} (2020) 062005}
  [\href{https://arxiv.org/abs/arXiv:2008.06486}{{\ttfamily
  arXiv:2008.06486}}].

\bibitem{2019ICRC...36..280H}
W.~{Hanlon}, \emph{{Telescope Array} 10 year composition},  in \emph{36th
  International Cosmic Ray Conference (ICRC2019)}, vol.~36 of
  \emph{International Cosmic Ray Conference}, p.~280, July, 2019
  [\href{https://arxiv.org/abs/arXiv:1908.01356}{{\ttfamily
  arXiv:1908.01356}}].

\bibitem{2019ICRC...36..482Y}
A.~{Yushkov}, \emph{Mass composition of cosmic rays with energies above
  $10^{17.2}$~{eV} from the hybrid data of the {Pierre Auger Observatory}},  in
  \emph{36th International Cosmic Ray Conference (ICRC2019)}, vol.~36 of
  \emph{International Cosmic Ray Conference}, p.~482, July, 2019,
  \href{https://doi.org/10.22323/1.358.0482}{DOI}.

\bibitem{TelescopeArray:2021gxg}
{\scshape Telescope Array, Pierre Auger} collaboration, \emph{{UHECR} arrival
  directions in the latest data from the original {Auger} and {TA} surface
  detectors and nearby galaxies},  in \emph{Proceedings of 37th International
  Cosmic Ray Conference {\textemdash} PoS(ICRC2021)}, vol.~395, p.~308, 2021,
  \href{https://doi.org/10.22323/1.395.0308}{DOI}.

\bibitem{Benson-Linsley-1981}
R.~{Benson} and J.~{Linsley}, \emph{Satellite observation of cosmic ray air
  showers},  in \emph{17th International Cosmic Ray Conference, Paris, France},
  vol.~8, pp.~145--148, 1981.

\bibitem{Bunner1967}
A.N.~{Bunner}, \emph{Cosmic Ray Detection by Atmospheric Fluorescence}, Ph.D.
  thesis, Cornell University, Jan., 1967.

\bibitem{1977ICRC....8..252M}
G.W.~{Mason}, H.E.~{Bergeson}, G.L.~{Cassiday}, T.W.~{Chiu}, D.A.~{Cooper},
  J.W.~{Elbert} et~al., \emph{Observations of extensive air showers by air
  fluorescence description of experimental techniques},  in \emph{International
  Cosmic Ray Conference}, vol.~8 of \emph{International Cosmic Ray Conference},
  p.~252, Jan., 1977.

\bibitem{1983AIPC...96..191C}
R.~{Cady}, G.L.~{Cassiday}, J.~{Elbert}, E.~{Loh}, Y.~{Mizumoto}, P.~{Sokolsky}
  et~al., \emph{{The Fly's Eye}},  in \emph{Science Underground}, vol.~96 of
  \emph{American Institute of Physics Conference Series}, pp.~191--202, Mar.,
  1983, \href{https://doi.org/10.1063/1.33936}{DOI}.

\bibitem{OWL}
J.F.~{Ormes}, L.M.~{Barbier}, K.~{Boyce}, E.~{Christian}, J.F.~{Krizmanic},
  J.F.~{Mitchell} et~al., \emph{Orbiting {Wide-angle Light Collectors} ({OWL}):
  A pair of {Earth} orbiting ``eyes'' to study air showers initiated by
  $>10^{20}$~{eV} particles},  in \emph{International Cosmic Ray Conference},
  vol.~5 of \emph{International Cosmic Ray Conference}, p.~273, Jan., 1997.

\bibitem{2001AIPC..566...57K}
B.A.~{Khrenov}, M.I.~{Panasyuk}, V.V.~{Alexandrov}, D.I.~{Bugrov},
  A.~{Cordero}, G.K.~{Garipov} et~al., \emph{Space program {KOSMOTEPETL}
  (project {KLYPVE} and {TUS}) for the study of extremely high energy cosmic
  rays},  in \emph{Observing Ultrahigh Energy Cosmic Rays from Space and
  Earth}, H.~{Salazar}, L.~{Villasenor} and A.~{Zepeda}, eds., vol.~566 of
  \emph{American Institute of Physics Conference Series}, pp.~57--75, May,
  2001, \href{https://doi.org/10.1063/1.1378622}{DOI}.

\bibitem{2001ICRC....2..831A}
V.V.~{Alexandrov}, D.I.~{Bugrov}, G.K.~{Garipov}, V.M.~{Grebenyuk},
  M.~{Finger}, B.A.~{Khrenov} et~al., \emph{Space experiment {TUS} for study of
  ultra high energy cosmic rays}, {\emph{International Cosmic Ray Conference}
  {\bfseries 2} (2001) 831}.

\bibitem{JEM-EUSO}
J.H.~{Adams}, S.~{Ahmad}, J.N.~{Albert}, D.~{Allard}, L.~{Anchordoqui},
  V.~{Andreev} et~al., \emph{The {JEM-EUSO} mission: An introduction},
  \href{https://doi.org/10.1007/s10686-015-9482-x}{\emph{Experimental
  Astronomy} {\bfseries 40} (2015) 3}.

\bibitem{POEMMA:2020ykm}
{\scshape POEMMA} collaboration, \emph{{The POEMMA (Probe of Extreme
  Multi-Messenger Astrophysics) observatory}},
  \href{https://doi.org/10.1088/1475-7516/2021/06/007}{\emph{Journal of
  Cosmology and Astroparticle Physics} {\bfseries 06} (2021) 007}
  [\href{https://arxiv.org/abs/arXiv:2012.07945}{{\ttfamily
  arXiv:2012.07945}}].

\bibitem{SSR2017}
P.A.~{Klimov}, M.I.~{Panasyuk}, B.A.~{Khrenov}, G.K.~{Garipov},
  N.N.~{Kalmykov}, V.L.~{Petrov} et~al., \emph{The {TUS} detector of extreme
  energy cosmic rays on board the {Lomonosov} satellite},
  \href{https://doi.org/10.1007/s11214-017-0403-3}{\emph{Space Science Reviews}
  {\bfseries 212} (2017) 1687}
  [\href{https://arxiv.org/abs/arXiv:1706.04976}{{\ttfamily
  arXiv:1706.04976}}].

\bibitem{JCAP2017}
B.A.~{Khrenov}, P.A.~{Klimov}, M.I.~{Panasyuk}, S.A.~{Sharakin},
  L.G.~{Tkachev}, M.Y.~{Zotov} et~al., \emph{First results from the {TUS}
  orbital detector in the extensive air shower mode},
  \href{https://doi.org/10.1088/1475-7516/2017/09/006}{\emph{Journal of
  Cosmology and Astroparticle Physics} {\bfseries 9} (2017) 006}
  [\href{https://arxiv.org/abs/arXiv:1704.07704}{{\ttfamily
  arXiv:1704.07704}}].

\bibitem{2019RemS...11.2449K}
P.~{Klimov}, B.~{Khrenov}, M.~{Kaznacheeva}, G.~{Garipov}, M.~{Panasyuk},
  V.~{Petrov} et~al., \emph{Remote sensing of the atmosphere by the ultraviolet
  detector {TUS} onboard the {Lomonosov} satellite},
  \href{https://doi.org/10.3390/rs11202449}{\emph{Remote Sensing} {\bfseries
  11} (2019) 2449}.

\bibitem{nuclearites-ICRC2019}
K.~{Shinozaki}, A.~{Montanaro}, M.E.~{Bertaina}, F.~{Fenu}, S.~{Ferrarese},
  P.~{Klimov} et~al., \emph{Search for nuclearites by the satellite-based {TUS}
  air fluorescence detector},  in \emph{36th International Cosmic Ray
  Conference (ICRC2019)}, vol.~36 of \emph{International Cosmic Ray
  Conference}, p.~545, July, 2019,
  \href{https://doi.org/10.22323/1.358.0545}{DOI}.

\bibitem{JCAP2020}
B.A.~{Khrenov}, G.K.~{Garipov}, M.A.~{Kaznacheeva}, P.A.~{Klimov},
  M.I.~{Panasyuk}, V.L.~{Petrov} et~al., \emph{An extensive-air-shower-like
  event registered with the {TUS} orbital detector},
  \href{https://doi.org/10.1088/1475-7516/2020/03/033}{\emph{Journal of
  Cosmology and Astroparticle Physics} {\bfseries 2020} (2020) 033}
  [\href{https://arxiv.org/abs/arXiv:1907.06028}{{\ttfamily
  arXiv:1907.06028}}].

\bibitem{Klimov:2021rxn}
{\scshape Lomonosov-UHECR/TLE} collaboration, \emph{Main results of the {TUS}
  experiment on board the {Lomonosov} satellite},  in \emph{Proceedings of 37th
  International Cosmic Ray Conference {\textemdash} PoS(ICRC2021)}, vol.~395,
  p.~316, 2021, \href{https://doi.org/10.22323/1.395.0316}{DOI}
  [\href{https://arxiv.org/abs/arXiv:2108.07022}{{\ttfamily
  arXiv:2108.07022}}].

\bibitem{CAPEL20182954}
F.~Capel, A.~Belov, M.~Casolino and P.~Klimov, \emph{{Mini-EUSO}: A high
  resolution detector for the study of terrestrial and cosmic {UV} emission
  from the {International Space Station}},
  \href{https://doi.org/https://doi.org/10.1016/j.asr.2017.08.030}{\emph{Advances
  in Space Research} {\bfseries 62} (2018) 2954}
  [\href{https://arxiv.org/abs/arXiv:1709.00405}{{\ttfamily
  arXiv:1709.00405}}].

\bibitem{Mini-EUSO2021}
S.~{Bacholle}, P.~{Barrillon}, M.~{Battisti}, A.~{Belov}, M.~{Bertaina},
  F.~{Bisconti} et~al., \emph{Mini-{EUSO} mission to study earth {UV} emissions
  on board the {ISS}},
  \href{https://doi.org/10.3847/1538-4365/abd93d}{\emph{Astrophysical Journal.
  Supplement Series} {\bfseries 253} (2021) 36}
  [\href{https://arxiv.org/abs/arXiv:2010.01937}{{\ttfamily
  arXiv:2010.01937}}].

\bibitem{Bertaina:2021+i}
{\scshape JEM-EUSO} collaboration, \emph{An overview of the {JEM-EUSO} program
  and results},  in \emph{Proceedings of 37th International Cosmic Ray
  Conference {\textemdash} PoS(ICRC2021)}, vol.~395, p.~406, 2021,
  \href{https://doi.org/10.22323/1.395.0406}{DOI}.

\bibitem{Panasyuk:20164y}
M.~Panasyuk, P.~Klimov, B.~Khrenov, S.~Sharakin, M.~Zotov, P.~Picozza et~al.,
  \emph{Ultra high energy cosmic ray detector {KLYPVE} on board the {Russian
  Segment} of the {ISS}},  in \emph{Proceedings of The 34th International
  Cosmic Ray Conference {\textemdash} PoS(ICRC2015)}, vol.~236, p.~669, 2016,
  \href{https://doi.org/10.22323/1.236.0669}{DOI}.

\bibitem{Klimov2017ICRC}
P.~Klimov and M.~Casolino, \emph{Status of the {KLYPVE-EUSO} detector for
  {EECR} study on board the {ISS}},  in \emph{35th International Cosmic Ray
  Conference (ICRC2017)}, vol.~301 of \emph{International Cosmic Ray
  Conference}, p.~412, 2017, \href{https://doi.org/10.22323/1.301.0412}{DOI}.

\bibitem{Bertaina:2019+0}
{\scshape JEM-EUSO} collaboration, \emph{Search for ultra-high energy cosmic
  rays from space~--- the {JEM-EUSO} program},  in \emph{Proceedings of 36th
  International Cosmic Ray Conference {\textemdash} PoS(ICRC2019)}, vol.~358,
  p.~192, 2019, \href{https://doi.org/10.22323/1.358.0192}{DOI}.

\bibitem{JE-exposure}
{\scshape JEM-EUSO} collaboration, \emph{{JEM-EUSO} observational technique and
  exposure},
  \href{https://doi.org/10.1007/s10686-014-9376-3}{\emph{Experimental
  Astronomy} {\bfseries 40} (2015) 117}.

\bibitem{2014ApJ...790L..21A}
R.U.~{Abbasi}, M.~{Abe}, T.~{Abu-Zayyad}, M.~{Allen}, R.~{Anderson}, R.~{Azuma}
  et~al., \emph{Indications of intermediate-scale anisotropy of cosmic rays
  with energy greater than 57 {EeV} in the {N}orthern sky measured with the
  surface detector of the {T}elescope {A}rray experiment},
  \href{https://doi.org/10.1088/2041-8205/790/2/L21}{\emph{Astrophysical
  Journal Letters} {\bfseries 790} (2014) L21}
  [\href{https://arxiv.org/abs/arXiv:1404.5890}{{\ttfamily arXiv:1404.5890}}].

\bibitem{Auger-dipole-2017}
{\scshape Pierre Auger} collaboration, \emph{{Observation of a large-scale
  anisotropy in the arrival directions of cosmic rays above
  $8{\times}10^{18}$~eV}},
  \href{https://doi.org/10.1126/science.aan4338}{\emph{Science} {\bfseries 357}
  (2017) 1266} [\href{https://arxiv.org/abs/arXiv:1709.07321}{{\ttfamily
  arXiv:1709.07321}}].

\bibitem{diMatteo:20197z}
{\scshape Pierre Auger, Telescope Array} collaboration, \emph{Full-sky searches
  for anisotropies in {UHECR} arrival directions with the {Pierre Auger
  Observatory} and the {Telescope Array}},  in \emph{Proceedings of 36th
  International Cosmic Ray Conference {\textemdash} PoS(ICRC2019)}, vol.~358,
  p.~439, 2019, \href{https://doi.org/10.22323/1.358.0439}{DOI}.

\bibitem{Kim:2021mcf}
{\scshape Telescope Array} collaboration, \emph{Hotspot update, and a new
  excess of events on the sky seen by the {Telescope Array} experiment},  in
  \emph{Proceedings of 37th International Cosmic Ray Conference {\textemdash}
  PoS(ICRC2021)}, vol.~395, p.~328, 2021,
  \href{https://doi.org/10.22323/1.395.0328}{DOI}.

\bibitem{2018ApJLett_starburst}
{\scshape Pierre Auger} collaboration, \emph{An indication of anisotropy in
  arrival directions of ultra-high-energy cosmic rays through comparison to the
  flux pattern of extragalactic gamma-ray sources},
  \href{https://doi.org/10.3847/2041-8213/aaa66d}{\emph{The Astrophysical
  Journal} {\bfseries 853} (2018) L29}
  [\href{https://arxiv.org/abs/arXiv:1801.06160}{{\ttfamily
  arXiv:1801.06160}}].

\bibitem{Tinyakov:2021}
P.~Tinyakov, L.~Anchordoqui, T.~Bister, J.~Biteau, L.~Caccianiga, R.~de~Almeida
  et~al., \emph{The {UHECR} dipole and quadrupole in the latest data from the
  original {Auger} and {TA} surface detectors},  in \emph{Proceedings of 37th
  International Cosmic Ray Conference {\textemdash} PoS(ICRC2021)}, vol.~395,
  p.~375, 2021, \href{https://doi.org/10.22323/1.395.0375}{DOI}.

\bibitem{Kalashev:2019skq}
O.~Kalashev, M.~Pshirkov and M.~Zotov, \emph{Identifying nearby sources of
  ultra-high-energy cosmic rays with deep learning},
  \href{https://doi.org/10.1088/1475-7516/2020/11/005}{\emph{Journal of
  Cosmology and Astroparticle Physics} {\bfseries 11} (2020) 005}
  [\href{https://arxiv.org/abs/arXiv:1912.00625}{{\ttfamily
  arXiv:1912.00625}}].

\bibitem{2017PTEP.2017lA101D}
B.R.~{Dawson}, M.~{Fukushima} and P.~{Sokolsky}, \emph{Past, present and future
  of {UHECR} observations},
  \href{https://doi.org/10.1093/ptep/ptx054}{\emph{Progress of Theoretical and
  Experimental Physics} {\bfseries 2017} (2017) }
  [\href{https://arxiv.org/abs/arXiv:1703.07897}{{\ttfamily
  arXiv:1703.07897}}].

\bibitem{2019FrASS...6...23B}
R.~{Alves Batista}, J.~{Biteau}, M.~{Bustamante}, K.~{Dolag}, R.~{Engel},
  K.~{Fang} et~al., \emph{Open questions in cosmic-ray research at ultrahigh
  energies}, \href{https://doi.org/10.3389/fspas.2019.00023}{\emph{Frontiers in
  Astronomy and Space Sciences} {\bfseries 6} (2019) 23}
  [\href{https://arxiv.org/abs/arXiv:1903.06714}{{\ttfamily
  arXiv:1903.06714}}].

\bibitem{jemeuso-meteors}
J.~Adams, J.H., S.~Ahmad, J.-N.~Albert, D.~Allard, L.~Anchordoqui, V.~Andreev
  et~al., \emph{{JEM-EUSO}: {Meteor} and nuclearite observations},
  \href{https://doi.org/10.1007/s10686-014-9375-4}{\emph{Experimental
  Astronomy} {\bfseries 40} (2015) 253}.

\bibitem{Piotrowski:2021a1}
{\scshape JEM-EUSO} collaboration, \emph{Towards observations of nuclearites in
  {Mini-EUSO}},  in \emph{Proceedings of 37th International Cosmic Ray
  Conference {\textemdash} PoS(ICRC2021)}, vol.~395, p.~503, 2021,
  \href{https://doi.org/10.22323/1.395.0503}{DOI}.

\bibitem{Paul:2021W0}
T.C.~Paul, S.T.~Reese, L.A.~Anchordoqui and A.V.~Olinto, \emph{{EUSO-SPB2}
  sensitivity to macroscopic dark matter},  in \emph{Proceedings of 37th
  International Cosmic Ray Conference {\textemdash} PoS(ICRC2021)}, vol.~395,
  p.~519, 2021, \href{https://doi.org/10.22323/1.395.0519}{DOI}.

\bibitem{Khrenov2004}
B.A.~{Khrenov}, V.V.~{Alexandrov}, D.I.~{Bugrov}, G.K.~{Garipov},
  N.N.~{Kalmykov}, M.I.~{Panasyuk} et~al., \emph{{KLYPVE/TUS space experiments
  for study of ultrahigh-energy cosmic rays}},
  \href{https://doi.org/10.1134/1.1825529}{\emph{Physics of Atomic Nuclei}
  {\bfseries 67} (2004) 2058}.

\bibitem{Garipov2001}
G.K.~Garipov, V.V.~Alexandrov, D.I.~Bugrov, A.~Cordero, M.~Cuautle,
  B.A.~Khrenov et~al., \emph{Electronics for the {KLYPVE} detector},
  \href{https://doi.org/10.1063/1.1378623}{\emph{AIP Conference Proceedings}
  {\bfseries 566} (2001) 76}.

\bibitem{Garipov2015BRAS}
G.K.~{Garipov}, M.Y.~{Zotov}, P.A.~{Klimov}, M.I.~{Panasyuk}, O.A.~{Saprykin},
  L.G.~{Tkachev} et~al., \emph{{The KLYPVE ultrahigh energy cosmic ray detector
  on board the ISS}},
  \href{https://doi.org/10.3103/S1062873815030193}{\emph{Bulletin of the
  Russian Academy of Sciences, Physics} {\bfseries 79} (2015) 326}.

\bibitem{JEM-EUSO-Instrument2015ExA}
J.H.~{Adams}, S.~{Ahmad}, J.N.~{Albert}, D.~{Allard}, L.~{Anchordoqui},
  V.~{Andreev} et~al., \emph{{The JEM-EUSO instrument}},
  \href{https://doi.org/10.1007/s10686-014-9418-x}{\emph{Experimental
  Astronomy} {\bfseries 40} (2015) 19}.

\bibitem{Adams2015-2}
J.H.~{Adams}, S.~{Ahmad}, J.N.~{Albert}, D.~{Allard}, L.~{Anchordoqui},
  V.~{Andreev} et~al., \emph{The {EUSO-Balloon} pathfinder},
  \href{https://doi.org/10.1007/s10686-015-9467-9}{\emph{Experimental
  Astronomy} {\bfseries 40} (2015) 281}.

\bibitem{Wiencke2017}
L.~Wiencke and A.~Olinto, \emph{{EUSO-SPB1} mission and science},  in
  \emph{Proceedings of 35th International Cosmic Ray Conference {\textemdash}
  PoS(ICRC2017)}, vol.~301, p.~1097, 2017,
  \href{https://doi.org/10.22323/1.301.1097}{DOI}.

\bibitem{Berat2010}
C.~{Berat}, S.~{Bottai}, D.~{De Marco}, S.~{Moreggia}, D.~{Naumov},
  M.~{Pallavicini} et~al., \emph{Full simulation of space-based extensive air
  showers detectors with {ESAF}},
  \href{https://doi.org/10.1016/j.astropartphys.2010.02.005}{\emph{Astroparticle
  Physics} {\bfseries 33} (2010) 221}
  [\href{https://arxiv.org/abs/arXiv:0907.5275}{{\ttfamily arXiv:0907.5275}}].

\bibitem{Fenu:2019+0}
{\scshape JEM-EUSO} collaboration, \emph{Simulations for the {JEM-EUSO} program
  with {ESAF}},  in \emph{Proceedings of 36th International Cosmic Ray
  Conference {\textemdash} PoS(ICRC2019)}, vol.~358, p.~252, 2019,
  \href{https://doi.org/10.22323/1.358.0252}{DOI}.

\bibitem{2021EUSO-SPB2_OptTest}
V.~Kungel, R.~Bachman, J.~Brewster, M.~Dawes, J.~Desiato, J.~Eser et~al.,
  \emph{{EUSO-SPB2} telescope optics and testing},  in \emph{Proceedings of
  37th International Cosmic Ray Conference {\textemdash} PoS(ICRC2021)},
  vol.~395, p.~412, 2021, \href{https://doi.org/10.22323/1.395.0412}{DOI}.

\bibitem{blin2014}
S.~Blin-Bondil, F.~Dulucq, J.~Rabanal, S.~Dagoret-Campagne, J.~Tongbong,
  P.~Barrillon et~al., \emph{{SPACIROC3}: {A} front-end readout {ASIC} for
  {JEM-EUSO} cosmic ray observatory},  in \emph{Proceedings of Technology and
  Instrumentation in Particle Physics 2014 {\textemdash} PoS(TIPP2014)},
  vol.~213, p.~172, 2015, \href{https://doi.org/10.22323/1.213.0172}{DOI}.

\bibitem{Belov2018}
A.A.~Belov, P.A.~Klimov and S.A.~Sharakin, \emph{The network architecture of
  the data-processing system for the photodetector of an orbital detector of
  ultra-high energy cosmic rays},
  \href{https://doi.org/10.1134/S0020441218010013}{\emph{Instruments and
  Experimental Techniques} {\bfseries 61} (2018) 27}.

\bibitem{Caibration2015ExA}
J.H.~{Adams}, S.~{Ahmad}, J.N.~{Albert}, D.~{Allard}, L.~{Anchordoqui},
  V.~{Andreev} et~al., \emph{{Calibration aspects of the {JEM-EUSO} mission}},
  \href{https://doi.org/10.1007/s10686-015-9453-2}{\emph{Experimental
  Astronomy} {\bfseries 40} (2015) 91}.

\bibitem{Fenu:2021wub}
{\scshape JEM-EUSO} collaboration, \emph{Expected performance of the {K-EUSO}
  space-based observatory},  in \emph{Proceedings of 37th International Cosmic
  Ray Conference {\textemdash} PoS(ICRC2021)}, vol.~395, p.~409, 2021,
  \href{https://doi.org/10.22323/1.395.0409}{DOI}.

\bibitem{2017NIMPA.866..150A}
G.~{Abdellaoui}, S.~{Abe}, A.~{Acheli}, J.H.~{Adams}, S.~{Ahmad}, A.~{Ahriche}
  et~al., \emph{Cosmic ray oriented performance studies for the {JEM-EUSO}
  first level trigger},
  \href{https://doi.org/10.1016/j.nima.2017.05.043}{\emph{Nuclear Instruments
  and Methods in Physics Research A} {\bfseries 866} (2017) 150}.

\bibitem{JEM-EUSO-exposure}
{\scshape JEM-EUSO} collaboration, \emph{An evaluation of the exposure in nadir
  observation of the {JEM-EUSO} mission},
  \href{https://doi.org/10.1016/j.astropartphys.2013.01.008}{\emph{Astroparticle
  Physics} {\bfseries 44} (2013) 76}
  [\href{https://arxiv.org/abs/arXiv:1305.2478}{{\ttfamily arXiv:1305.2478}}].

\bibitem{EA_Angle}
{\scshape JEM-EUSO} collaboration, \emph{Performances of {JEM-EUSO}: angular
  reconstruction},
  \href{https://doi.org/10.1007/s10686-013-9371-0}{\emph{Experimental
  Astronomy} {\bfseries 40} (2015) 153}.

\bibitem{EA_Energy}
{\scshape JEM-EUSO} collaboration, \emph{Performances of {JEM–EUSO}: energy
  and {X}$_\mathrm{max}$ reconstruction},
  \href{https://doi.org/10.1007/s10686-014-9427-9}{\emph{Experimental
  Astronomy} {\bfseries 40} (2015) 183}.

\bibitem{BMY2020}
M.Y.~{Zotov} and D.B.~{Sokolinskiy}, \emph{The first application of neural
  networks to the analysis of the {TUS} orbital detector data},
  \href{https://doi.org/10.3103/S0027134920060235}{\emph{Moscow University
  Physics Bulletin} {\bfseries 75} (2020) 657}.

\bibitem{2021Univ....7..221Z}
M.~{Zotov}, \emph{Application of neural networks to classification of data of
  the {TUS} orbital telescope},
  \href{https://doi.org/10.3390/universe7070221}{\emph{Universe} {\bfseries 7}
  (2021) 221} [\href{https://arxiv.org/abs/arXiv:2106.03361}{{\ttfamily
  arXiv:2106.03361}}].

\bibitem{2018APh....97...46E}
M.~{Erdmann}, J.~{Glombitza} and D.~{Walz}, \emph{A deep learning-based
  reconstruction of cosmic ray-induced air showers},
  \href{https://doi.org/10.1016/j.astropartphys.2017.10.006}{\emph{Astroparticle
  Physics} {\bfseries 97} (2018) 46}
  [\href{https://arxiv.org/abs/arXiv:1708.00647}{{\ttfamily
  arXiv:1708.00647}}].

\bibitem{Ivanov:2020nfo}
D.~Ivanov, O.E.~Kalashev, M.Y.~Kuznetsov, G.I.~Rubtsov, T.~Sako, Y.~Tsunesada
  et~al., \emph{Using deep learning to enhance event geometry reconstruction
  for the {Telescope Array} surface detector},
  \href{https://doi.org/10.1088/2632-2153/abae74}{\emph{Mach. Learn. Sci.
  Tech.} {\bfseries 2} (2021) 015006}
  [\href{https://arxiv.org/abs/arXiv:2005.07117}{{\ttfamily
  arXiv:2005.07117}}].

\bibitem{PierreAuger:2021fkf}
{\scshape Pierre Auger} collaboration, \emph{Deep-learning based reconstruction
  of the shower maximum {X}$_\mathrm{max}$ using the water-{Cherenkov}
  detectors of the {Pierre Auger Observatory}},
  \href{https://doi.org/10.1088/1748-0221/16/07/P07019}{\emph{JINST} {\bfseries
  16} (2021) P07019} [\href{https://arxiv.org/abs/arXiv:2101.02946}{{\ttfamily
  arXiv:2101.02946}}].

\bibitem{Filippatos:2021b5}
G.~Filippatos, M.~Battisti, M.E.~Bertaina, F.~Bisconti, J.~Eser, G.~Osteria
  et~al., \emph{Expected performance of the {EUSO-SPB2} fluorescence
  telescope},  in \emph{Proceedings of 37th International Cosmic Ray Conference
  {\textemdash} PoS(ICRC2021)}, vol.~395, p.~405, 2021,
  \href{https://doi.org/10.22323/1.395.0405}{DOI}.

\bibitem{Eser:2021mbp}
{\scshape JEM-EUSO} collaboration, \emph{Science and mission status of
  {EUSO-SPB2}},  in \emph{Proceedings of 37th International Cosmic Ray
  Conference {\textemdash} PoS(ICRC2021)}, vol.~395, p.~404, 2021,
  \href{https://doi.org/10.22323/1.395.0404}{DOI}.

\bibitem{Brummel:2013urn}
V.~Br\"ummel, R.~Engel and M.~Roth, \emph{On the importance of the energy
  resolution for identifying sources of {UHECR}},  in \emph{{33rd International
  Cosmic Ray Conference}}, p.~0667, 2013.

\bibitem{jem-euso-atmos}
{\scshape JEM-EUSO} collaboration, \emph{Science of atmospheric phenomena with
  {JEM-EUSO}},
  \href{https://doi.org/10.1007/s10686-014-9431-0}{\emph{Experimental
  Astronomy} {\bfseries 40} (2015) 239}.

\bibitem{Marcelli:2021uX}
L.~Marcelli, E.~Arnone, M.~Barghini, M.~Battisti, A.S.~Belov, M.E.~Bertaina
  et~al., \emph{Observation of {ELVES} with {Mini-EUSO} telescope on board the
  {International Space Station}},  in \emph{Proceedings of 37th International
  Cosmic Ray Conference {\textemdash} PoS(ICRC2021)}, vol.~395, p.~367, 2021,
  \href{https://doi.org/10.22323/1.395.0367}{DOI}.

\bibitem{Golzio:2021tv}
A.~Golzio, M.~Battisti, M.E.~Bertaina, K.~Bolmgren, G.~Cambiè, M.~Casolino
  et~al., \emph{A study on {UV} emission from clouds with {Mini-EUSO}},  in
  \emph{Proceedings of 37th International Cosmic Ray Conference {\textemdash}
  PoS(ICRC2021)}, vol.~395, p.~208, 2021,
  \href{https://doi.org/10.22323/1.395.0208}{DOI}.

\end{thebibliography}\endgroup
\end{document}